\documentclass[preprint,12pt,authoryear]{elsarticle}

\usepackage{amssymb,amsmath,natbib,scalefnt,tablefootnote,float}
\usepackage{amsthm,amssymb}
\usepackage{mathtools}
\usepackage{etoolbox}
\AtBeginEnvironment{align}{\setcounter{subeqn}{0}}
\newcounter{subeqn} \renewcommand{\thesubeqn}{\theequation\alph{subeqn}}%
\newcommand{\subeqn}{%
	\refstepcounter{subeqn}
	\tag{\thesubeqn}
}
\usepackage{mathrsfs}
\usepackage{footnote}
\usepackage{hyperref}
\usepackage{graphicx}
\usepackage{nomencl}
\usepackage{subcaption}
\usepackage{xcolor}

\definecolor{brght}{rgb}{0.825,0.2625,0.15}

\allowdisplaybreaks
\journal{Journal of the Mechanics and Physics of Solids}
\makeatletter
\def\ps@pprintTitle{%
	\let\@oddhead\@empty
	\let\@evenhead\@empty
	\let\@oddfoot\@empty
	\let\@evenfoot\@oddfoot
}
\makeatother
\begin{document}

\begin{frontmatter}



\title{Multiphysics analysis of acoustically actuated nanospherical antennas embedded in polymer/metal medium with magneto-electro-elastic surface/interface effects} 


\author[a]{Mohsen Farsiani} 
\author[{a,b}]{Hossein M. Shodja\corref{cor1}}
\cortext[cor1]{Corresponding author}
\emailauthor{shodja@sharif.edu}{Hossein M. Shodja}
\address[a]{Department of Civil Engineering, Sharif University of
	Technology, P.O. Box 11155-9313, Tehran, Iran}
\address[b]{Center for Nanoscience and Nanotechnology, Institute for Convergence Science \& Technology, Sharif University of Technology, Tehran 14588-89694, Iran}

\begin{abstract}
A precise analytical treatment for predicting the behavior of nano-sized magneto-electro-elastic (MEE) antennas and resonators under incident acoustic waves requires careful consideration of multiphysics surface/interface effects, including magnetization, polarization, and elasticity. To date, no analytical solutions have incorporated all three phenomena simultaneously. By addressing these surface effects, this work presents a rigorous mathematical analysis of a nano-sized spherically isotropic embedded MEE spherical shell subjected to incident acoustic waves. The set of coupled spectral constitutive relations relevant to the bulk of the MEE spherical shell is distinguished from those pertinent to its free inner surface and matrix-shell interface. The surrounding matrix may consist of an isotropic dielectric or metallic material. Conventional electrodynamics theories are insufficient to address this problem, as they do not adequately account for MEE effects at the surface or interface. To overcome this limitation, the study employs the equivalent impedance matrix (EIM) method combined with surface/interface elasticity to model the surface/interface MEE behaviors rigorously. For metallic matrices, a plasmonics-based mathematical framework is utilized, with the optical properties described by the plasma model to accurately capture metallic behavior. The spectral EIM method, combined with vector and tensor spherical harmonics forming a Schauder basis for square-integrable vector fields and second-rank symmetric tensor fields on the unit sphere, is shown to be a pivotal tool for solving the fully coupled elastodynamics and Maxwell's equations. This approach is particularly effective in capturing significant MEE surface/interface effects. This methodology enables a detailed exploration of surface/interface characteristic lengths, facilitating the examination of size-dependent effects on electromagnetic radiated power and fundamental resonance frequency. The findings provide valuable insights into the behavior of acoustically actuated nanospherical antennas, nanosensors, and nanoresonators based on MEE nanospheres. Moreover, these results have significant implications for the design and optimization of nanoscale devices in advanced technological applications.
\end{abstract}


\end{frontmatter}

\section{Introduction}
The current work investigates the antenna efficiency and resonance frequency of an embedded nano-sized magneto-electro-elastic (MEE) spherical shell, contrasting it with nano-sized piezoelectric spherical shell \cite{f50}, nano-sized piezoelectric fibers \cite{f2}, and nano-sized magneto-electric (ME) plates \cite{f20}. Hence, the development of an appropriate multiphysics theoretical method that provides a precise coupled-effects description of the magnetic, electric, and elastic phenomena associated with the free surface and interface of an acoustically excited embedded nanosized MEE spherical shell is of great interest. To this end, we introduce an innovative analytical approach that incorporates Gurtin-Murdoch (GM) surface/interface elasticity theory \cite{m29} in conjunction with the equivalent impedance matrix (EIM) method, based on transmission line theory \cite{f47,f49}. On one hand, the magnetic and electric performances of the free surface and interface are modeled via EIM; on the other hand, their elasticity effects are accounted for through GM theory. As an immediate application of the theoretical developments presented herein, the design of nanosized MEE antennas and resonators may be highlighted.

The growing need for rapid data transfer is pushing the boundaries of network bandwidths, often surpassing the limitations of current infrastructure. To address these challenges, terahertz (THz) frequencies have gained attention as a potential game-changer \cite{f17}. Antennas, which act as a crucial link between electric currents and electromagnetic (EM) waves in technologies like smartphones and RFID systems, play a key role in this evolution. The push for increasingly compact wireless devices has driven extensive research into shrinking components, particularly antennas, which are essential for transmitting and receiving EM signals \cite{f5}. However, the physical size of compact antennas often exceeds one-tenth of the EM wavelength, presenting significant challenges at high frequencies where wavelengths are relatively large \cite{f9,f10,f11,f12}. To meet the demands of next-generation wireless technologies, innovative antenna designs for THz applications are vital, balancing the need for extreme miniaturization with optimal performance.

Recent advancements have introduced magnetoelectric (ME) antennas that combine acoustic wave resonance with the magnetoelectric effect, showing potential in the very-high frequency and ultra-high frequency  bands. This innovative approach helps mitigate the efficiency and size challenges encountered by conventional electrically small antennas \cite{f13,f14,f15,f16,f20}. Unlike traditional antennas, which generate EM waves through oscillating charges, ME antennas exploit the oscillation of magnetic dipole moments. These oscillations are acoustically induced at the electromechanical resonance frequency, rather than the EM wave resonance frequency. Since acoustic waves propagate much slower than EM waves at the same operational frequency, this mechanism allows for a significant reduction in antenna size, achieving dimensions one to two orders of magnitude smaller than conventional designs \cite{f20}. 

The generalized spectral magneto-electro-elastodynamic interface theory presented in \cite{f50} is effective for analyzing piezoelectric particles embedded in a dielectric matrix, accounting for both surface/interface polarization and surface/interface elasticity effects; where the surface/interface magnetization effects were negligible.  For that purpose, GM theory which in its original form is an augmented elasticity theory was first extended to account for surface/interface polarization effects.  Therefore, it cannot be used to study problems involving MEE embedded particles. To address fully dynamic problems with nanosized MEE antennas embedded in dielectric or metallic matrices with magnetized interfaces, a more suitable method must be developed and applied. This challenge is one of the primary objectives of the present work.

An alternative model for magnetized surfaces/interfaces and the surrounding medium, including MEE materials, can be developed based on the analogy between plane-wave propagation in free space and signal transmission in a transmission line. In this analogy, the electric and magnetic fields of a propagating wave in free space correspond to the voltages and currents of a signal in the equivalent transmission line. The components of the EIM of the surface/interface serve as the macroscopic parameters of this system (e.g., \cite{f52}; \cite{f59}; \cite{f60} for non-mechanical cases).

In metallic matrices, the optical properties over a wide frequency range are explained by the plasma model, where a gas of free electrons with number density 
$n$ moves against a fixed background of positive ion cores \cite{f61}. The plasma model simplifies electron behavior by neglecting lattice potential and electron-electron interactions, instead assuming that certain aspects of the band structure are captured in the effective optical mass $m$ of each electron. Electrons respond to the applied electromagnetic field, and their motion is damped by collisions with a characteristic frequency $\gamma  = \frac{1}{\tau }$, where $\tau$ is the relaxation time of the free electron gas.

Building on these concepts, the present work investigates the behavior of an acoustically actuated nanospherical MEE particle embedded in a polymer/metal matrix and subjected to terahertz (THz) frequencies. We employ the EIM method to model the fully electromagnetic behavior of MEE media. For the metallic matrix, we apply a plasmonics-based mathematical framework to accurately capture its behavior. Given the spherical geometry of the MEE particle, a rigorous mathematical analysis is made possible through utilization of vector and tensor spherical harmonics \cite{m10,f42,f43,m16}, which form an orthogonal Schauder basis for the space of square-integrable functions on ${{\mathbb{S}}^2}$ (see e.g. \cite{f32}, Chapter 7).
\section{Model}
\begin{figure}[H]
	\begin{center}
		\includegraphics[width=12cm]{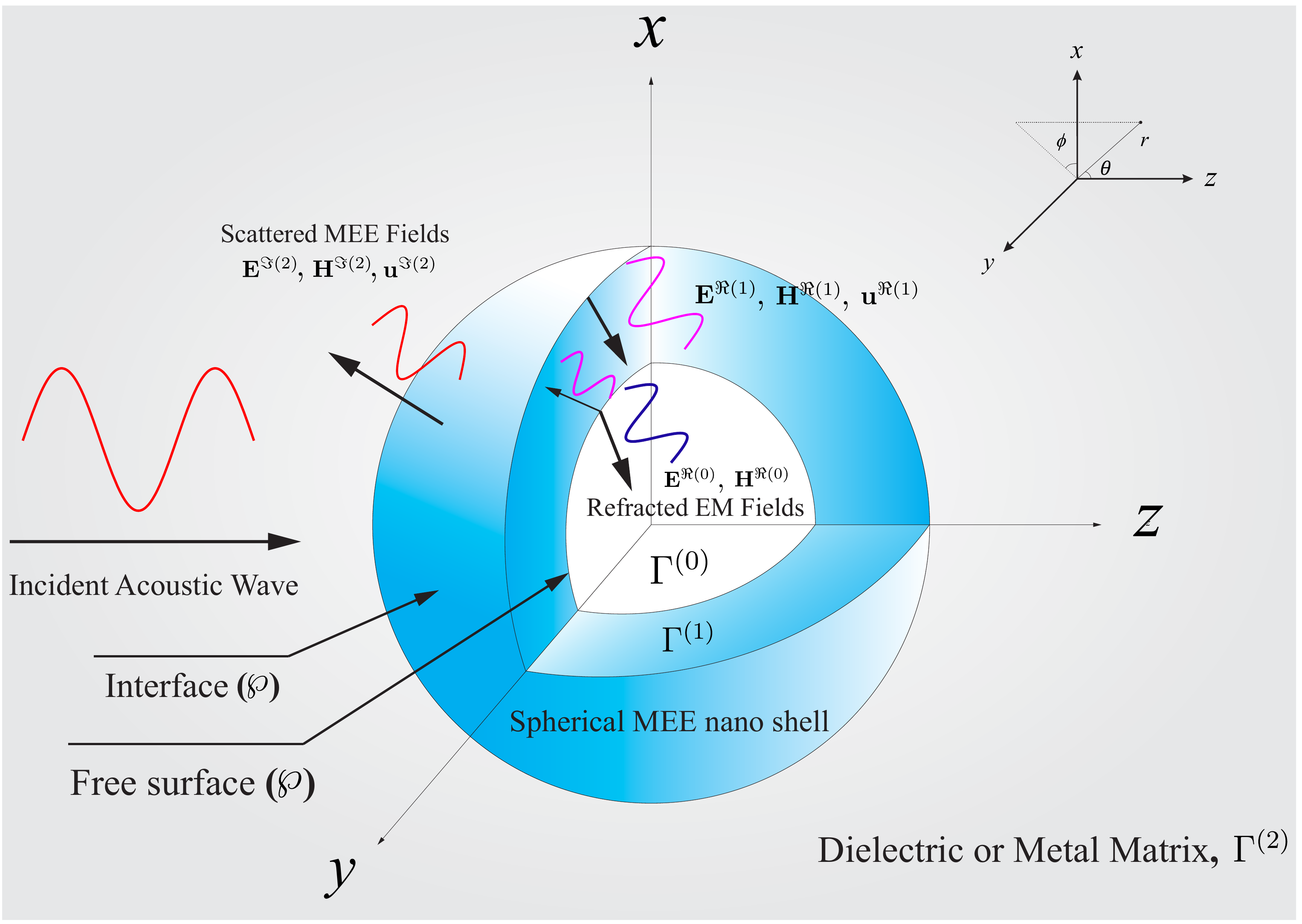}
	\end{center}
	\caption{Schematic of the nomenclature for the proposed embedded nano-sized MEE shell subjected to incident acoustic waves, emphasizing key parameters and variables.}
	\label{30000}
\end{figure}
Consider an isotropic and homogeneous dielectric/metal matrix, designated as ${\Gamma ^{(2)}}$, encasing a nanoscale magneto-electro-elastic (MEE) hollow sphere with inner radius ${R_1}$ and outer radius ${R_2}$, referred to as ${\Gamma ^{(1)}}$, as illustrated in Fig.~\ref{30000}. It's important to note that the MEE hollow sphere is made from a multiferroic material, showcasing a combination of ferromagnetic, ferroelectric, and ferroelastic properties. This material is synthesized as a composite, incorporating specific volume fractions of piezomagnetic and piezoelectric materials.  In the forthcoming section on numerical examples, we will delve into multiferroic materials comprising cobalt ferrite (CoFe\textsubscript{2}O\textsubscript{4}) and barium titanate (BaTi{O\textsubscript{3}}), which represent distinct types of piezomagnetic and piezoelectric materials, respectively. In this work, we address both dielectric (polymer) matrix and metal matrix scenarios, both of which involve the multiphysics of MEE fields. It is well-known that in the case of metal matrix the state of plasmonic phenomenon is realized. In the plasmonic state, the optical properties of metals are commonly explained by a Drude model, which demonstrates that these properties vary with the frequency of the propagating waves.The geometries and the domains occupied by the constituents of the multiphase medium of interest are depicted in Fig. 1; the details of Fig. 1 are summarized as below:
\begin{equation}\label{eq1}
\left\{ \begin{gathered}
r < {R_1} \hfill \qquad \text{spherical nanovoid}\hspace{0.25em}({\Gamma ^{(0)}})\\
r = {R_1} \hfill \qquad \text{nanovoid MEE surface}\hspace{0.25em}(\wp)\\
{R_1} < r < {R_2} \hfill \qquad \text{MEE spherical wall}\hspace{0.25em}({\Gamma ^{(1)}})\\
r = {R_2} \hfill \qquad \text{matrix-MEE sphere interface}\hspace{0.25em}(\wp)\\
r > {R_2} \hfill \qquad \text{dielectric/metal matrix}\hspace{0.25em}({\Gamma ^{(2)}})\\ 
\end{gathered}  \right.
\end{equation}
The presence of the superscript $\wp$ above a field quantity signifies its association with either the interface or the free surface. Additionally, the superscripts “$(0)$”, “$(1)$”, and “$(2)$” are used to denote the quantities pertinent to the spherical nanovoid, sphere wall, and the matrix regions, respectively.

The point of origin for both the Cartesian coordinate system $(x, y, z)$ and the spherical coordinate system $(r, \theta, \phi)$ is set at the center of the spherical shell, as illustrated in Figure~\ref{30000}. To streamline notation, hereafter $(\mathit{\Omega})$ is used to represent $(\theta, \phi)$. We make the assumption of perfect bonding and a coherent interface between the shell and the matrix. All the points within the wall of the embedded nanosized MEE hollow sphere are assumed to have spherically isotropic MEE properties with radial polarization. The particle is subjected to an incident high-frequency mechanical time-harmonic P-wave with an angular frequency $\omega$, propagating in the positive $z$-direction. Since the hollow sphere under investigation is nano-sized, in order to be detectable by the in coming P-wave, the frequency of the wave must be within the THz range. Another critical issue is that, at this scale the MEE hollow sphere has a very large surface-to-bulk ratio, and thus the nanoscopic physical and mechanical parameters pertinent to its inner and outer surfaces become important. More specifically, the nanoscopic MEE fields and the residual stresses (${\boldsymbol{\sigma}_0}$) within, respectively, the sphere interface with its surrounding dielectric/metal matrix and the sphere free inner surface are determining factors. In this study, one of the main objectives is to extend the mathematical theory of GM surface elasticity theory (\cite{m29}) to the field of electrodynamics theory to accommodate MEE surfaces and interfaces. To this end, the idea of the the equivalent impedance matrix and transmission line theory for modeling surfaces and interfaces will be utilized. The extended theory encounters certain boundary conditions which differ from the usuall classical boundary conditions. As it will be seen the new nonclassical MEE boundary conditions, in contrast to the classical Dirichlet and Neumann type boundary conditions, involve the second derivatives of certain MEE fields (displacement, electric field, and magnetic field). The task is to obtain an accurate analytical solution within both the dielectric/metal matrix and the embedded nano-sized MEE hollow sphere, governed by the fully dynamic and fully coupled Maxwell's and elastodynamics equations subjected to the noted nonclassical MEE boundary conditions. Particular attention will be given to the understanding of how the interface and the free surface of the MEE nano-sized hollow sphere affect the scattered MEE fields once it is impinged upon by the incoming acoustic P-waves.

We will tackle the task of solving the governing system of coupled MEE partial differential equations by utilizing scalar, vector, and tensor spherical harmonics (\cite{m10,f48}). The accuracy, efficacy, and high convergence rate of the solution with only a few harmonics have been illustrated in recent works (\cite{f43}).

In the absence of body forces, electric charges, and current densities within both the dielectric/metal matrix and the MEE hollow sphere, the fully coupled Maxwell's and elastodynamics equations describing the MEE scattered fields are:
\begin{equation}
{\textrm{div}}_{{{\mathbb{R}^3}}}\boldsymbol{\bar \sigma}  = \rho \frac{{{\partial ^2}\mathbf{\bar u}}}{{{\partial ^2}t}},
\end{equation}
in ${\Gamma ^{(1)}} \cup {\Gamma ^{(2)}}$, and
\begin{subequations}\label{eq10000}
	\begin{align}
	&{\textrm{curl}}_{{{\mathbb{R}^3}}} \mathbf{\bar E} =  - \frac{{\partial \mathbf{\bar B}}}{{\partial t}},\label{eq2000} \\
	&{\textrm{curl}}_{{{\mathbb{R}^3}}} \mathbf{\bar H} = \frac{{\partial \mathbf{\bar D}}}{{\partial t}},\\
	&{\textrm{div}}_{{{\mathbb{R}^3}}}\mathbf{\bar D} = 0,\\
	&{\textrm{div}}_{{{\mathbb{R}^3}}}\mathbf{\bar B} = 0,\label{eq2004}
	\end{align}
\end{subequations}
in ${\Gamma ^{(0)}} \cup {\Gamma ^{(1)}} \cup {\Gamma ^{(2)}}$.\\
In the equations above, the symbols $\rho$, $\boldsymbol{\bar \sigma}$, $\mathbf{\bar u}$, $\mathbf{\bar E}$, $\mathbf{\bar H}$, $\mathbf{\bar D}$, and $\mathbf{\bar B}$ represent the mass density, stress field, displacement field, electric field, magnetic field, electric displacement, and magnetic flux density, respectively. We are interested in solutions where the time variation of all the MEE field variables, ($\boldsymbol{\bar \sigma}$, $\mathbf{\bar u}$, $\mathbf{\bar E}$, $\mathbf{\bar H}$, $\mathbf{\bar D}$, and $\mathbf{\bar B}$) are assumed to be time harmonic with angular frequency $\omega$, as
\begin{align}
&\{ \mathbf{\bar u}, \boldsymbol{\bar \sigma}, \mathbf{\bar E}, \mathbf{\bar H}, \mathbf{\bar D}, \mathbf{\bar B}\} (r,\mathit{\Omega} ,t)\notag\\
&\qquad = \{ \mathbf{u}, \boldsymbol{\sigma}, \mathbf{E}, \mathbf{H}, \mathbf{D}, \mathbf{B}\} (r,\mathit{\Omega} ){e^{ - \iota \omega t}}\label{eq2001}
\end{align}
with $\iota  = \sqrt { - 1}$. In the above relation $\{ \mathbf{u}, \boldsymbol{\sigma}, \mathbf{E}, \mathbf{H}, \mathbf{D}, \mathbf{B}\} (r,\mathit{\Omega} )$ are the amplitudes of their corresponding field quantities.

Consequently, Eqs.~ \eqref{eq10000} and \eqref{eq2000}-\eqref{eq2004} leads to:
\begin{equation}\label{eq5000}
{\textrm{div}}_{{{\mathbb{R}^3}}}\boldsymbol{\sigma}  + \rho {\omega ^2}\mathbf{u} = 0,
\end{equation}
in ${\Gamma ^{(1)}} \cup {\Gamma ^{(2)}}$, and
\begin{subequations}
	\begin{align}
	&{\textrm{curl}}_{{{\mathbb{R}^3}}} \mathbf{E} = \iota \omega \mathbf{B}, \label{eq1000}\\
	&{\textrm{curl}}_{{{\mathbb{R}^3}}} \mathbf{H} =  - \iota \omega \mathbf{D},  \label{eq1001}\\
	&{\textrm{div}}_{{{\mathbb{R}^3}}}\mathbf{D} = 0,  \label{eq200}\\
	&{\textrm{div}}_{{{\mathbb{R}^3}}}\mathbf{B} = 0,  \label{eq201}
	\end{align}
\end{subequations}
in ${\Gamma ^{(0)}} \cup {\Gamma ^{(1)}} \cup {\Gamma ^{(2)}}$.\\
It is worth noting that Eqs.\eqref{eq200} and \eqref{eq201} can be derived by applying the divergence operator to Eqs.\eqref{eq1001} and \eqref{eq1000}, respectively.

Under small strain theory assumption, the strain-displacement relation is:
\begin{equation}\label{eq998}
\boldsymbol{\epsilon}  = \frac{1}{2}\big({\nabla _{{\mathbb{R}^3}}} \mathbf{u} + ({{\nabla _{{\mathbb{R}^3}}} \mathbf{u})^T}\big),
\end{equation}\\
where $\boldsymbol{\epsilon}$ is the strain tensor.

We use the superscripts "$\Re$", "$\mathcal{I}$", or "$\Im$" over a field quantity to signify that the quantity corresponds to the refracted, incident, or scattered fields, respectively. Consequently, the MEE fields may be expressed as
\begin{subequations}\label{eq11}
	\begin{align}
	&\mathbf{u} = \left\{ \begin{array}{l}
	{\mathbf{u}^{\Re(1)}},{R_1}  <  r < {R_2}\\
	{\mathbf{u}^{\mathcal{I}(2)}} + {\mathbf{u}^{\Im(2)}},r > {R_2}
	\end{array}, \right.\label{eq40}\\
	&\mathbf{E} = \left\{ \begin{array}{l}
	{\mathbf{E}^{\Re(0)}},r < {R_1}\\
	{\mathbf{E}^{\Re(1)}},{R_1} < r < {R_2}\\
	{\mathbf{E}^{\Im(2)}},r > {R_2}
	\end{array}, \right.\\
	&\mathbf{H} = \left\{ \begin{array}{l}
	{\mathbf{H}^{\Re(0)}},r < {R_1}\\
	{\mathbf{H}^{\Re(1)}},{R_1} < r < {R_2}\\
	{\mathbf{H}^{\Im(2)}},r > {R_2}
	\end{array}.\right.
	\end{align}
\end{subequations}
In this study, we postulate that the displacement field arising from the incident P-wave is described as follows:
\begin{subequations}
	\begin{align}
	&u_x^{\mathcal{I}(2)}(x,y,z,t) = 0,\\
	&u_y^{\mathcal{I}(2)}(x,y,z,t) = 0,\\
	&u_z^{\mathcal{I}(2)}(x,y,z,t) = {\Lambda}{e^{\iota ({K_p}z - \omega t)}},
	\end{align}
\end{subequations}
In essence, we assume that the P-wave takes the form of a harmonic plane wave characterized by an angular frequency denoted as $\omega$ and an amplitude represented by $\Lambda$ along the $z$-direction. In this context, $K_p$ signifies the compressive wave number, expressed as
\begin{equation}
{K_p} = \frac{\omega }{{{C_p}}},
\end{equation}
Here, ${C_p} = \sqrt {\frac{{\lambda + 2\mu }}{\rho^{(2)}}}$ represents the velocity of the compressive wave, with $\lambda$ and $\mu$ denoting the Lame' constants of the dielectric/metal matrix. As we delve into the subsequent discussions, it will become apparent that expressing the constitutive equations of the spherically isotropic MEE region becomes more convenient when utilizing spherical coordinates $(r,\theta ,\phi )$. To facilitate this, let's denote the standard orthonormal basis associated with spherical coordinates $(r,\theta ,\phi )$ as ${\mathbf{e}_r}$, ${\mathbf{e}_\theta }$, and ${\mathbf{e}_\phi }$. Subsequently, the MEE fields can be expanded in the following manner:
\begin{equation}
\left\{ \begin{gathered}
\mathbf{u} = {u_i}{\mathbf{e}_i}, \hfill \\
\boldsymbol{\epsilon}  = {\epsilon _{ij}}{\mathbf{e}_i} \otimes {\mathbf{e}_j}, \hfill \\
\boldsymbol{\sigma}  = {\sigma _{ij}}{\mathbf{e}_i} \otimes {\mathbf{e}_j}, \hfill \\
\end{gathered}  \right.
\qquad \text{in}\hspace{0.25em}{\Gamma ^{(1)}} \cup {\Gamma ^{(2)}}
\end{equation}
\begin{align}
&\{ \mathbf{E},\mathbf{D},\mathbf{H},\mathbf{B}\}  = \{ {E_i},{D_i},{H_i},{B_i}\} {\mathbf{e}_i}, \\
&\quad\text{in}\hspace{0.25em}{\Gamma ^{(0)}} \cup{\Gamma ^{(1)}} \cup {\Gamma ^{(2)}}.\notag
\end{align}
Within this framework, we designate $i$ and $j$ as the indices representing the directions $r$, $\theta$, and $\phi$. To formulate the constitutive relations, we will employ the subsequent notation to represent the components of the elastic moduli tensor $\mathbf{C}$, the piezoelectric tensor $\mathbf{e}$, piezomagnetic tensor $\boldsymbol{\gamma}$, the dielectric tensor $\boldsymbol{\kappa}$, the magneto-electric tensor $\mathbf{\mathscr{G}}$, magnetic permeability tensor $\boldsymbol{\mu}$:
\begin{align}
&{C_{11}} = {C_{22}} = {C_{\theta \theta \theta \theta }} = {C_{\phi \phi \phi \phi }},\quad{C_{12}} = {C_{\theta \theta \phi \phi }},\notag\\
&{C_{13}} = {C_{23}} = {C_{\theta \theta rr}} = {C_{\phi \phi rr}},\quad{C_{33}} = {C_{rrrr}},\notag\\
&{C_{44}} = {C_{r\phi r\phi }} = {C_{r\theta r\theta }},\quad{e_{31}} = {e_{32}} = {e_{r\theta \theta }} = {e_{r\phi \phi }},\label{eq1002}\\
&{e_{33}} = {e_{rrr}},\quad{e_{15}} = {e_{\phi \phi r}} = {e_{\theta \theta r}},\notag\\
&{\gamma_{31}} = {\gamma_{32}} = {\gamma_{r\theta \theta }} = {\gamma_{r\phi \phi }},\quad{\gamma_{33}} = {\gamma_{rrr}},\notag\\
&{\gamma_{15}} = {\gamma_{\phi \phi r}} = {\gamma_{\theta \theta r}},\notag\\
&{\kappa_{11}} = {\kappa_{22}} = {\kappa_{\theta \theta }} = {\kappa_{\phi \phi }},\quad{\kappa_{33}} = {\kappa_{rr}},\notag\\
&{\mathscr{G}_{11}} = {\mathscr{G}_{22}} = {\mathscr{G}_{\theta \theta }} = {\mathscr{G}_{\phi \phi }},\quad{\mathscr{G}_{33}} = {\mathscr{G}_{rr}},\notag\\
&{\mu_{11}} = {\mu_{22}} = {\mu_{\theta \theta }} = {\mu_{\phi \phi }},\quad{\mu_{33}} = {\mu_{rr}}.\notag
\end{align}
Subsequently, the constitutive equations for the spherically isotropic MEE shell can be expressed conveniently in the following manner:
\begin{equation}\label{eq3}
\begin{split}
&{\sigma _{\theta \theta }} = {C_{11}}{\epsilon _{\theta \theta }} + {C_{12}}{\epsilon _{\phi \phi }} + {C_{13}}{\epsilon _{rr}} - {e_{31}}{E_r}- {\gamma_{31}}{H_r},\\
&{\sigma _{\phi \phi }} = {C_{12}}{\epsilon _{\theta \theta }} + {C_{11}}{\epsilon _{\phi \phi }} + {C_{13}}{\epsilon _{rr}} - {e_{31}}{E_r}- {\gamma_{31}}{H_r},\\
&{\sigma _{rr}} = {C_{13}}{\epsilon _{\theta \theta }} + {C_{13}}{\epsilon _{\phi \phi }} + {C_{33}}{\epsilon _{rr}} - {e_{33}}{E_r}- {\gamma_{33}}{H_r},\\
&{\sigma _{r\theta }} = 2{C_{44}}{\epsilon _{r\theta }} - {e_{15}}{E_\theta }- {\gamma_{15}}{H_\theta },\\
&{\sigma _{r\phi }} = 2{C_{44}}{\epsilon _{r\phi }} - {e_{15}}{E_\phi }- {\gamma_{15}}{H_\phi },\\
&{\sigma _{\theta \phi }} = \big({C_{11}} - {C_{12}}\big){\epsilon _{\theta \phi }},\\
&{D_\theta } = 2{e_{15}}{\epsilon _{r\theta }} + {\kappa_{11}}{E_\theta }+ {\mathbf{\mathscr{G}}_{11}}{H_\theta },\\
&{D_\phi } = 2{e_{15}}{\epsilon _{r\phi }} + {\kappa_{11}}{E_\phi }+ {\mathbf{\mathscr{G}}_{11}}{H_\phi },\\
&{D_r} = {e_{31}}{\epsilon _{\theta \theta }} + {e_{31}}{\epsilon _{\phi \phi }} + {e_{33}}{\epsilon _{rr}} + {\kappa_{33}}{E_r}+ {\mathbf{\mathscr{G}}_{33}}{H_r},\\
&{B_\theta } = 2{\gamma_{15}}{\epsilon _{r\theta }} + {\mu_{11}}{H_\theta }+ {\mathbf{\mathscr{G}}_{11}}{E_\theta },\\
&{B_\phi } = 2{\gamma_{15}}{\epsilon _{r\phi }} + {\mu_{11}}{H_\phi }+ {\mathbf{\mathscr{G}}_{11}}{E_\phi },\\
&{B_r} = {\gamma_{31}}{\epsilon _{\theta \theta }} + {\gamma_{31}}{\epsilon _{\phi \phi }} + {\gamma_{33}}{\epsilon _{rr}} + {\mu_{33}}{H_r}+ {\mathbf{\mathscr{G}}_{33}}{E_r},
\end{split}
\end{equation}
in ${\Gamma ^{(1)}}$. It should be emphasized that the MEE constitutive equations mentioned above, pertaining to the considered hollow spherical multiferroic particle, incorporate the coupled phenomena of ferromagnetism, ferroelectricity, and ferroelasticity encountered in the problem at hand.

Additionally, we can express the constitutive relationships for the elastic dielectric/metal matrix as follows:
\begin{equation}\label{eq997}
\begin{split}
&{\sigma _{ij}} = 2\mu {\epsilon _{ij}} + \lambda tr(\boldsymbol{\epsilon} ){\delta _{ij}},\\
&{D_i} = {\kappa_m} {E_i},\\
&{B_i} = {\mu _m}{H_i},
\end{split}
\end{equation}
within ${\Gamma ^{(2)}}$. In Eqs.~\eqref{eq997}, $i$ and $j$ represent $r$, $\theta$, and $\phi$, the symbols $\lambda$ and $\mu$ denote the Lame' constants, and ${\kappa_m}$ and ${\mu _m}$ represent, respectively, the dielectric function and the magnetic permeability of the matrix. The dielectric function is defined as:
\begin{equation}\label{eq234}
{\kappa _m}=\left\{ \begin{gathered}
\kappa _m^d\quad\quad\textrm{for dielectric matrix,}\hfill\\
{\kappa _0} (1- \frac{{\omega _p^2}}{{{\omega ^2}}})\quad\quad\textrm{for metal matrix,}\hfill\\ 
\end{gathered}  \right.
\end{equation}
and:
\begin{equation}
\omega _p^2 = \frac{{n{e^2}}}{{{\kappa _0}m}},
\end{equation}
where $n$ represents the number density of electrons, $e$ denotes the electric charge, $\kappa_0$ is the vacuum electric permitivity, and $m$ stands for the effective mass of the electron.

From Drude's model which is represented by Eq.~\eqref{eq234} it is clearly seen that the dielectric function of the free electron plasma for the metal matrix is dependent upon the propagating wave frequency.
\section{MEE boundary conditions presented within surface elasticity theory and utilizing the equivalent impedance matrix method}\label{Sec3}
In view of the fact that the embedded MEE hollow sphere is on the nano-scale, the role of its free inner surface and its interface with the surrounding matrix have a significant effect on the generated MEE fields.  Given the circumstances, the resolution of the accuracy of the treatment of the encountered coupled multiphysics phenomenon is increased by a two-fold analysis.  The mechanical effects of the interface and the free surface are accounted for through incorporation of GM surface/interface elasticity theory \cite{f50}, while the magnetic and the electric performance of the free surface/interface are incorporated through consideration of the equivalent impedance matrix (EIM) method which is based on the idea of the transmission line theory \cite{f47,f49}.  Assuming that the surface/interface displacement field, ${\mathbf{{\bar u}}^\wp }(\mathit{\Omega}, t)$ and the surface/interface stress field ${\boldsymbol{{\bar \sigma}}^\wp }(\mathit{\Omega}, t)$ are time harmonic, then ${\mathbf{{\bar u}}^\wp }(\mathit{\Omega}, t)={\mathbf{u}^\wp }{e^{ - \iota \omega t}}$ and ${\boldsymbol{{\bar \sigma}}^\wp }(\mathit{\Omega}, t)={\boldsymbol{\sigma}^\wp }{e^{ - \iota \omega t}}$.  Consequently, the elastodynamic equation of motion corresponding to such a surface/interface with unit outward normal vector $\mathbf{n}$ may be written as:
\begin{equation}\label{eq2002}
\textrm{div}_{\wp }{\boldsymbol{\sigma} ^\wp } + [\boldsymbol{\sigma} .\mathbf{n}] = - {\rho ^\wp }{\omega ^2}{\mathbf{u}^\wp } \qquad \text{on}\hspace{0.25em}{\wp}.
\end{equation}
Here, ${\rho ^\wp }$ denotes the mass density at the interface, and $[\boldsymbol{\sigma} .\mathbf{n}]$ represents the jump in the quantity $\boldsymbol{\sigma} .\mathbf{n}$ across the surface/interface.\\
Assuming perfect interface bonding between the dielectric/metal matrix - hollow MEE sphere, we have:
\begin{equation}\label{eq20000}
[\mathbf{u}] = 0\qquad \text{on}\hspace{0.25em}{r={R_2}}.
\end{equation}
Suppose that the surface/interface magnetization and polarization can be expressed as ${\mathbf{M} ^\wp }(\mathit{\Omega}, t)={\mathbf{M}^\wp }{e^{ - \iota \omega t}}$ and ${\mathbf{P} ^\wp }(\mathit{\Omega}, t)={\mathbf{P}^\wp }{e^{ - \iota \omega t}}$, respectively.  Recently, by accounting for the surface/interface magnetization and polarization, the generalization of GM theory to MEE surfaces/interfaces lead to (\cite{f50}):
\begin{subequations}
	\begin{align}
	&\mathbf{n} \times [\mathbf{E}]  - \iota \omega {\mu _0}{\mathbf{M}^\wp}= 0 \quad\hspace{0.19em}\hspace{0.25em} \text{on}\hspace{0.25em}r={R_1},\hspace{0.25em} \text{and}\hspace{0.25em}r={R_2}\label{eq415}\\
	&\mathbf{n} \times [\mathbf{H}] + \iota \omega {\mathbf{P}^\wp} = 0 \qquad\hspace{0.25em}\hspace{0.25em} \text{on}\hspace{0.25em}r={R_1},\hspace{0.25em} \text{and}\hspace{0.25em}r={R_2} \label{eq416}
	\end{align}
\end{subequations}
Furthermore, we take inspiration from the work of Eringen (as presented in \cite{f51}) and the method of Gibbs dividing surface to craft the comprehensive constitutive equations for the MEE surface/interface, which are detailed as follows:
\begin{align}
&{\boldsymbol{\sigma} ^\wp } = {\boldsymbol{\sigma}_0} + {\mathbf{C}^\wp }:{\boldsymbol{\epsilon} ^\wp } - {({\mathbf{e}^\wp })^T}\cdot{\mathbf{E}^\wp } - {({\boldsymbol{\gamma}^\wp })^T}\cdot{\mathbf{H}^\wp },\hfill \label{eq2003}\\
&{\mathbf{P}^\wp } = {\mathbf{P}_0} + {\mathbf{e}^\wp }:{\boldsymbol{\epsilon} ^\wp } + {\boldsymbol{\kappa} ^\wp }\cdot{\mathbf{E}^\wp }+{\boldsymbol{\mathscr{G}} ^\wp }\cdot{\mathbf{H}^\wp },\hfill\label{eq2005}\\
&{\mathbf{M}^\wp } = {\mathbf{M}_0} + {\boldsymbol{\gamma}^\wp }:{\boldsymbol{\epsilon} ^\wp } + {\boldsymbol{\mathscr{G}} ^\wp }\cdot{\mathbf{E}^\wp }+{\boldsymbol{\mu} ^\wp }\cdot{\mathbf{H}^\wp },\hfill\label{eq2006}
\end{align}
in which $\mathbf{M}_0$ is the residual surface / interface magnetization tensor (with the units of [A]), $\mathbf{P}_0$ is the residual surface / interface electric polarization vector (with the units of $[\mathrm{C}/{\mathrm{m}}]$), $\boldsymbol{\sigma}_0$ represents the residual surface / interface stress tensor (with the units of $[\mathrm{N}/{\mathrm{m}}]$), $\mathbf{C}^\wp$ is the surface / interface elastic moduli tensor (with the units of $[\mathrm{N}/{\mathrm{m}}]$), $\mathbf{e}^\wp$ is the surface / interface piezoelectric tensor (with the units of $[\mathrm{C}/{\mathrm{m}}]$), $\boldsymbol{\gamma}^\wp$ is the surface / interface piezomagnetic tensor (with the units of $[\mathrm{N}/{\mathrm{A}}]$), $\boldsymbol{\kappa}^\wp$ is the surface / interface electric polarizability tensor (with the units of $[{\mathrm{C}^2}/\mathrm{Nm}]$), $\boldsymbol{\mathscr{G}}^\wp$ is the surface / interface magnetoelectric tensor (with the units of [Nsm/VC]), and $\boldsymbol{\mu}^\wp$ is the surface / interface magnetizability tensor (with the units of $[\mathrm{N}{\mathrm{s}^2}\mathrm{m}/{\mathrm{C}^2}]$).  Moreover, $\boldsymbol{\epsilon}^\wp$ is the surface / interface strain tensor and can be calculated as below:
\begin{equation}\label{eq419}
{\boldsymbol{\epsilon} ^\wp} = \frac{1}{2}\bigg((\mathcal{D}{\mathbf{u}^{\wp t}}) + {(\mathcal{D}{\mathbf{u}^{\wp t}})^T}\bigg) - {u^{\wp n}}\mathbf{L},
\end{equation}
where ${u^{\wp n}}$ and ${\mathbf{u}^{\wp t}}$ represent the normal and tangential components of the surface / interface displacement vector, respectively.  The operator $\mathbf{L}$ is the Weingarten map of the surface, $\mathcal{D}{\mathbf{u}^{\wp t}}$ is the tangential part of ${\nabla _\wp }{\mathbf{u}^{\wp t}}$ where ${\nabla _\wp }$ is the surface gradient (\cite{f50}).

It should be recalled that in the current study the nanosized spherical particle is made of a spherically isotropic MEE material which is embedded within an isotropic dielectric/metal matrix.  Therefore, the surface/interface constitutive relations given by Eqs. \eqref{eq2003} - \eqref{eq2006} for the problem of interest reduce to the following equations:
\begin{subequations}
	\begin{align}
	&\sigma _{\theta \theta }^\wp = {({\boldsymbol{\sigma} _0})_{\theta \theta }}+C_{11}^\wp\epsilon _{\theta \theta }^\wp + C_{12}^\wp\varepsilon _{\phi \phi }^\wp, \hfill \label{eq370}\\
	&\sigma _{\phi \phi }^\wp = {({\boldsymbol{\sigma} _0})_{\phi \phi }}+C_{12}^\wp\varepsilon _{\theta \theta }^\wp + C_{11}^\wp\varepsilon _{\phi \phi }^\wp,\hfill \\
	&\sigma _{\theta \phi }^\wp = {({\boldsymbol{\sigma} _0})_{\theta \phi }}+(C_{11}^\wp - C_{12}^\wp)\varepsilon _{\theta \phi }^\wp,\hfill \\
	&P_\theta ^\wp = {({\mathbf{P}_0})_\theta }+\kappa _{11}^\wp{E_\theta }+\mathscr{G} _{11}^\wp{H_\theta }, \hfill \label{eq371} \\
	&P_\phi ^\wp = {({\mathbf{P}_0})_\phi }+\kappa _{11}^\wp{E_\phi }+\mathscr{G} _{11}^\wp{H_\phi }, \hfill \label{eq372}\\
	&M_\theta ^\wp = {({\mathbf{M}_0})_\theta }+\mathscr{G} _{11}^\wp{E_\theta }+\mu _{11}^\wp{H_\theta }, \hfill \\
	&M_\phi ^\wp = {({\mathbf{M}_0})_\phi }+\mathscr{G} _{11}^\wp{E_\phi }+\mu _{11}^\wp{H_\phi }. \hfill \label{eq373}
	\end{align}
\end{subequations}
In contrast, the recent investigation in \cite{f50} examines a scenario wherein the region ${\Gamma ^{(1)}}$ is composed of a piezoelectric material rather than a MEE material. More strictly speaking, the study in \cite{f50} focuses on a spherical piezoelectric shell and neglects the surface/interface magnetization. It should be emphasized that in the current study, the spherical region with a central cavity is made of a MEE material and thus, in general, the surface/interface magnetization is is not zero, representing a departure from the previous assumption. To recap, the coupled governing equations provided by Eq.\eqref{eq5000} and Eqs.\eqref{eq1000} - \eqref{eq201} are to be solved subjected to the nonstandard surface/interface boundary conditions given by Eqs. \eqref{eq2002}, \eqref{eq20000}, \eqref{eq415}, and \eqref{eq416}.  For an efficient treatment of the surface / interface electro-magnetization, we propose to employ EIM method which is discussed in the next section.
\subsection{Emergence of the nonstandard surface/interface magnetoelectric boundary conditions through the EIM method}\label{2-1}
In view of the surface/interface MEE constitutive Eqs.~\eqref{eq370} - \eqref{eq373} for a spherical particle with MEE properties, it is evident that the mechanical and magnetoelectric fields within the surface / interface are decoupled.  Given the circumstances, a coherent framework for modeling the magnetoelectric behavior of MEE surfaces/interfaces can be established by drawing parallels between the propagation of plane waves and the transmission of signals within a transmission line.  This can be achieved by employing the EIM method; see, for example, \cite{f52} and \cite{f53}. In this context, the electric and magnetic fields of a traveling wave can, in effect, be made equivalent to the voltages and currents of a signal traveling through an equivalent transmission line. Additionally, the elements of the EIM of the surface/interface assume the function of its macroscopic properties.

\begin{figure}[H]
	\begin{center}
		\includegraphics[width=10cm]{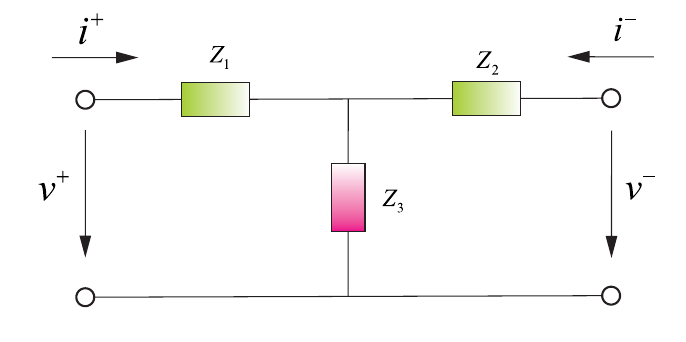}
	\end{center}
	\caption{The equivalent T circuit for the surface/interface ME boundary conditions.}
	\label{100}
\end{figure}

The field discontinuities in the left sides of Eqs.\eqref{eq415} and \eqref{eq416} can be referred to as effective total electric and magnetic currents, i.e., $\mathbf{n} \times [{\mathbf{H}_t^ +}-{\mathbf{H}_t^ -} ]={\mathbf{J}_{tot,e}}$ and ${\mathbf{E}_t^ +}-{\mathbf{E}_t^ -} =\mathbf{n} \times{\mathbf{J}_{tot,m}}$. 
Hence, these effective currents are expressed as:
\begin{subequations}
	\begin{align}
	&{\mathbf{J}_{tot,e}} =  - \iota \omega {\mathbf{P}^\wp },\\
	&{\mathbf{J}_{tot,m}} =  - \iota \omega {\mu _0}{\mathbf{M}^\wp }.
	\end{align}
\end{subequations}
By establishing a connection between the electric fields ${\mathbf{E}_t^ +}$ and ${\mathbf{E}_t^ -}$ and the voltages $v^+$ and $v^-$, along with the magnetic fields $\mathbf{n} \times {\mathbf{H}_t^ +}$ and $\mathbf{n} \times {\mathbf{H}_t^ -}$ and the currents $i^+$ and $-i^-$, while taking into consideration that $\mathbf{n} \times [{\mathbf{H}_t^ +}-{\mathbf{H}t^ -} ]={\mathbf{J}_{tot,e}}$ and ${\mathbf{E}_t^ +}-{\mathbf{E}_t^ -} =\mathbf{n} \times{\mathbf{J}_{tot,m}}$, a surface/interface of this nature can be represented using the so-called "equivalent T-circuit" (or sometimes referred to as "equivalent $\Pi$-circuit"); see, for example, \cite{f53}. The equivalent T-circuit transmission line representing the surface / interface boundary conditions encountered in the current study is depicted in Figure~\ref{100}. In this figure $Z_1$, $Z_2$, and $Z_3$ represent the impedances, characterizing the surface / interface ME behavior. Accordingly, the voltages on both sides of the surface/interface boundaries can be expressed with respect to the currents as follows:
\begin{align}
&{v^ + } = ({Z_1} + {Z_3}){i^ + } + {Z_3}{i^ - },\\
&{v^ - } = {Z_3}{i^ + } + ({Z_2} + {Z_3}){i^ - }.
\end{align}
More conveniently, the above equations are written in the following matrix form as:
\begin{equation}\label{eq1610}
\left( \begin{array}{l}
{v^ + }\\
{v^ - }
\end{array} \right) = \left( {\begin{array}{*{20}{c}}
	{{Z_{11}}}&{{Z_{12}}}\\
	{{Z_{21}}}&{{Z_{22}}}
	\end{array}} \right)\cdot\left( \begin{array}{l}
{i^ + }\\
{i^ - }
\end{array} \right)
\end{equation}
in which $Z_{11} = Z_1 + Z_3$, $Z_{12} = Z_{21} = Z_3$, and $Z_{22} = Z_2 + Z_3$. The coefficient matrix presented in Eq.~\eqref{eq1610} is referred to as the surface's/interface's EIM. Thus, the effective tangential ME fields at both sides of the surface/interface boundaries of the hollow embedded spherical MEE particle under consideration, by analogy with the transmission-line model, can be represented by the following expression:
\begin{equation}\label{eq1611}
\left( \begin{array}{l}
{\mathbf{E}_t^ + }\\
{\mathbf{E}_t^ - }
\end{array} \right) = \left( {\begin{array}{*{20}{c}}
	{{Z_{11}}}&{{Z_{12}}}\\
	{{Z_{12}}}&{{Z_{22}}}
	\end{array}} \right)\left( \begin{array}{l}
{\mathbf{n} \times \mathbf{H}_t^ +}\\
{-\mathbf{n} \times \mathbf{H}_t^ - }
\end{array} \right).
\end{equation}
Given the ME properties of the surface/interface boundaries of the MEE particle of interest or equivalently given the elements of the EIM, then the surface/interface boundary conditions \eqref{eq415} and \eqref{eq416} are equivalently replaced by Eq.~\eqref{eq1611}. In summary, the governing MEE equations \eqref{eq5000}, and \eqref{eq1000} - \eqref{eq201} must be solved subjected to the nonstandard boundary condition \eqref{eq2002}, perfect interface condition \eqref{eq20000}, and equivalent ME boundary condition \eqref{eq1611}.

The representation of an equivalent impedance matrix is a highly effective tool for modeling surfaces or interfaces of diverse complexity. It offers numerous advantages. Firstly, when the fields on both sides of the surface or interface are specified, it becomes straightforward to compute the impedances as defined in Eq.~\eqref{eq1611}. Secondly, with this information, one can determine various characteristics, such as whether the surface or interface should be capacitive (when $\operatorname{Im} [{Z_{ij}}] < 0$) or inductive (when $\operatorname{Im} [{Z_{ij}}] > 0$), whether it exhibits lossiness (when $\operatorname{Re} [{Z_{ij}}] > 0$) or activity (when $\operatorname{Re} [{Z_{ij}}] < 0$), whether it adheres to reciprocity ($Z_{12} = Z_{21}$) or is non-reciprocal ($Z_{12}  \ne Z_{21}$), whether it is symmetrical ($Z_{11} = Z_{22}$) or asymmetrical ($Z_{11} \ne Z_{22}$), and so forth.
Assuming reciprocity and symmetrical EIM, the ME surface/interface boundary conditions become:
\begin{subequations}
	\begin{align}
	&\left( \begin{array}{l}
	{\mathbf{E}_\theta^ + }\\
	{\mathbf{E}_\theta^ - }
	\end{array} \right) = \left( {\begin{array}{*{20}{c}}
		{{Z_{11}}}&{{Z_{12}}}\\
		{{Z_{12}}}&{{Z_{11}}}
		\end{array}} \right)\cdot\left( \begin{array}{l}
	{- \mathbf{H}_\phi^ +}\\
	{ \mathbf{H}_\phi^ - }
	\end{array} \right)\hfill\label{eq422}\\& \qquad \text{on}\hspace{0.25em}r={R_1},\hspace{0.25em} \text{and}\hspace{0.25em}r={R_2}\notag\\\notag\\
	&\left( \begin{array}{l}
	{\mathbf{E}_\phi^ + }\\
	{\mathbf{E}_\phi^ - }
	\end{array} \right) = \left( {\begin{array}{*{20}{c}}
		{{Z_{11}}}&{{Z_{12}}}\\
		{{Z_{12}}}&{{Z_{11}}}
		\end{array}} \right)\cdot\left( \begin{array}{l}
	{\mathbf{H}_\theta^ +}\\
	{ -\mathbf{H}_\theta^ - }
	\end{array} \right)\hfill\label{eq423}\\& \qquad \text{on}\hspace{0.25em}r={R_1},\hspace{0.25em} \text{and}\hspace{0.25em}r={R_2}\notag
	\end{align}
\end{subequations}
As shown in Appendix \ref{App new}, in the special case where an ultra-thin shell made of a piezoelectric material is embedded in an isotropic matrix and exhibits neither bulk magnetization nor surface and interface magnetization, the impedance values simplify to ${Z_{11}}={Z_{12}}=\frac{\iota }{{\omega {\kappa ^\wp }}}$. 

It should be noted that along any arbitrary path on the surface the surface traction, surface polarization, and the surface magnetization are tangent to the surface, and thus the components of these field quantities which are not tangent to the noted path are zero:
\begin{equation}
\sigma _{rr}^\wp  = \sigma _{r\theta }^\wp  = \sigma _{r\phi }^\wp  = P_r^\wp  = M_r^\wp  = 0
\end{equation}
\section{Spectral analysis of the MEE fields of a multiferroic nano-sized particle subjected to P-waves: the phenomenon of plasmonics}\label{Sec4}
Our main strategy in analyzing the problem under consideration is to transform the coupled governing boundary value problem (BVP), which includes tensor partial differential equations (PDEs) (Eq.\eqref{eq5000} and Eqs.\eqref{eq1000} - \eqref{eq201}, and \eqref{eq2002}, \eqref{eq20000}, \eqref{eq422}, \eqref{eq423}), into a set of ordinary differential equations (ODEs). This can be accomplished by expressing the unknown fields in the equations using spherical harmonics. Scalar spherical harmonics, denoted by ${\left\{ {{Y^{l,m}}(\Omega )} \right\}_{\substack{{l \in {\mathbb{N}_0}}\\m \leqslant l}}}$, are functions on the unit sphere ${{\mathbb{S}}^2}$ in $\mathbb{R}^3$. Each ${Y^{l,m}}(\Omega )$ is defined as ${P_{l,m}}({\mathop{\rm cos}\nolimits} \hspace{.2em}\theta ){e^{\iota m\phi }}$, where ${P_{l,m}}(z)$ is the associated Legendre polynomial of degree $l$ and order $m$. These functions are eigenfunctions of the surface Laplacian, given by $- \nabla _{{\mathbb{S}}^2}^2{Y^{l,m}}(\Omega ) = l(l + 1){Y^{l,m}}(\Omega )$. Notably, they form an orthogonal Schauder basis for the space of square-integrable functions on ${{\mathbb{S}}^2}$ (see e.g. \cite{f32}, Chapter 7).

In \cite{m27}, vectorial and tensorial spherical harmonics are introduced as bases for square-integrable vector and second-rank symmetric tensor fields on ${{\mathbb{S}}^2}$, respectively. Formulas from \cite{m28}, with some normalization constants omitted, are as follows:\\
Vector spherical harmonics: ($l \in {\mathbb{N}_0}$, $\left| m \right| \le l$)
\begin{subequations}
	\begin{align}
	&\mathbf{V}_1^{l,m}(\mathit{\Omega} ) = {Y^{l,m}}(\mathit{\Omega} ){\mathbf{e}_r},\label{eq7-1}\\
	&\mathbf{V}_2^{l,m}(\mathit{\Omega} ) = {\boldsymbol{\nabla} _{{\mathbb{S}}^2}}{Y^{l,m}}(\mathit{\Omega} ),\label{eq7-2}\\
	&\mathbf{V}_3^{l,m}(\mathit{\Omega} ) = {\mathbf{e}_r} \times {\boldsymbol{\nabla} _{{\mathbb{S}}^2}}{Y^{l,m}}(\mathit{\Omega} )\label{eq7-3}.
	\end{align}
\end{subequations}
Note that as $\mathbf{e}_r$ is normal to the sphere and $\mathbf{V}_2^{l,m}(\mathit{\Omega} )$ is tangent to the surface, $\mathbf{V}_3^{l,m}(\mathit{\Omega} ) = {\mathbf{e}_r} \times {\boldsymbol{\nabla} _{{\mathbb{S}}^2}}{Y^{l,m}}(\mathit{\Omega})$ will also be tangent to the sphere.\\
Second-rank tensor spherical harmonics: (${l \in {\mathbb{N}_0}}$ and $\left| m \right| \le l$)
\begin{subequations}
	\begin{align}
	&\mathbf{T}_1^{l,m}(\mathit{\Omega} ) = {Y^{l,m}}(\mathit{\Omega} ){\mathbf{e}_r} \otimes {\mathbf{e}_r},\label{eq9-1}\\
	&\mathbf{T}_2^{l,m}(\mathit{\Omega} ) = {Y^{l,m}}(\mathit{\Omega} ){\mathbf{e}_\theta } \otimes {\mathbf{e}_\theta } + {Y^{l,m}}(\mathit{\Omega} ){\mathbf{e}_\phi } \otimes {\mathbf{e}_\phi },\label{eq9-2}\\
	&\mathbf{T}_3^{l,m}(\mathit{\Omega} ) = 2{[{\mathbf{e}_r} \otimes {\boldsymbol{\nabla} _{{\mathbb{S}}^2}}{Y^{l,m}}(\mathit{\Omega} )]^S},\label{eq9-3}\\
	&\mathbf{T}_4^{l,m}(\mathit{\Omega} ) = {[{\boldsymbol{\nabla} _{{\mathbb{S}}^2}}{\boldsymbol{\nabla} _{{\mathbb{S}}^2}}{Y^{l,m}}(\mathit{\Omega} )]^{STT}},\label{eq9-4}\\
	&\mathbf{T}_5^{l,m}(\mathit{\Omega} ) = 2{[{\mathbf{e}_r} \otimes ({\mathbf{e}_r} \times {\boldsymbol{\nabla} _{{\mathbb{S}}^2}}{Y^{l,m}}(\mathit{\Omega} ))]^S},\label{eq9-5}\\
	&\mathbf{T}_6^{l,m}(\mathit{\Omega} ) =  - {[{\mathbf{e}_r} \times {\boldsymbol{\nabla} _{{\mathbb{S}}^2}}{\boldsymbol{\nabla} _{{\mathbb{S}}^2}}{Y^{l,m}}(\mathit{\Omega} )]^{STT}}.\label{eq9-6}
	\end{align}
\end{subequations}
In Eqs.~\eqref{eq9-1}-\eqref{eq9-6}, ${\left[ {...} \right]^S}$ represents the symmetric part of the quantity within the brackets, while ${\left[ {...} \right]^{TT}}$ denotes the transverse traceless part of the second-order tensor $\left[ {...} \right]$, as defined by:
\begin{equation}\label{eq}
{[{\mathscr{T}_{ab}}]^{TT}} = {\mathscr{P}_{aj}}{\mathscr{P}_{bk}}{\mathscr{T}_{jk}} - \frac{1}{2}{\mathscr{P}_{ab}}({\mathscr{P}_{jk}}{\mathscr{T}_{kj}}),
\end{equation}
where ${\mathscr{P}_{jk}} = {\delta _{jk}} - {n_j}{n_k}$, $m,n = 1,2,3$, ${\delta _{mn}}$ represents the Kronecker delta function, and $\mathbf{n} = {\mathbf{e}_r}$ is the unit radial vector (see, e.g., \cite{f3}).

To reduce the governing PDEs (Eq. \eqref{eq5000} and Eqs. \eqref{eq1000} - \eqref{eq201}) which describe the coupled MEE fields of the problem of interest to a system of ODEs, the MEE fields associated with both the bulk region and the surface/interface boundaries are expanded in terms of the spherical harmonics as follows:
\begingroup
\begin{align}
& \mathbf{u}(r;\mathit{\Omega} )  = \sum\limits_{l,m} {\sum\limits_{i = 1}^3 { u_i^{l,m}(r) \mathbf{V}_i^{l,m}(\mathit{\Omega} )} }\label{eq2013},\\
&{\rm{\{ \mathbf{D}}}\left( {r;\mathit{\Omega} } \right),{\rm{\mathbf{E}}}\left( {r;\mathit{\Omega} } \right)\}= \mathop \sum \limits_{l,m} \mathop \sum \limits_{i = 1}^3 \left\{ {D_i^{l,m}\left( r \right),E_i^{l,m}\left( r \right)} \right\}\mathbf{V}_i^{l,m}\left( \mathit{\Omega}  \right),\label{eq2016}\\
&{\rm{\{ \mathbf{B}}}\left( {r;\mathit{\Omega} } \right),{\rm{\mathbf{H}}}\left( {r;\mathit{\Omega} } \right)\}= \mathop \sum \limits_{l,m} \mathop \sum \limits_{i = 1}^3 \left\{ {B_i^{l,m}\left( r \right),H_i^{l,m}\left( r \right)} \right\}\mathbf{V}_i^{l,m}\left( \mathit{\Omega}  \right),\label{eq2017} \\
&\left\{ {\boldsymbol{\epsilon} (r;\Omega ),\boldsymbol{\sigma} (r;\Omega )} \right\}= \sum\limits_{l,m} {\sum\limits_{i = 1}^6 {\left\{ {\epsilon _i^{l,m}(r),\sigma _i^{l,m}(r)} \right\}\mathbf{T}_i^{l,m}(\mathit{\Omega} )} },\label{eq2014}\\
&{\rm{\{ {\mathbf{u}^\wp}}}\left( {\mathit{\Omega} } \right),{\mathbf{P}^\wp}\left( {\mathit{\Omega} } \right), {\mathbf{M}^\wp}\left( {\mathit{\Omega} } \right)\} \notag\\
& \qquad\qquad= \mathop \sum \limits_{l,m} \mathop \sum \limits_{i = 1}^3 \left\{ {u_i^{l,m;\wp},P_i^{l,m;\wp}, M_i^{l,m;\wp}} \right\}\mathbf{V}_i^{l,m}\left( \mathit{\Omega}  \right),\label{eq2015}\\
&\left\{ {{\boldsymbol{\epsilon}^\wp} (\mathit{\Omega} ),{\boldsymbol{\sigma}^\wp} (\mathit{\Omega} )} \right\}= \sum\limits_{l,m;\wp} {\sum\limits_{i = 1}^6 {\left\{ {\epsilon _i^{l,m;\wp},\sigma _i^{l,m;\wp}} \right\}\mathbf{T}_i^{l,m}(\mathit{\Omega} )} }.\label{eq2018}
\end{align}
\endgroup
Due to the rotational symmetry of the problem, the MEE fields are independent of $\phi$, leading to non-zero coefficients only for modes with $m=0$; it should be noted that, for $m=0$, ${P_{l,0}}$ represents the Legendre polynomial of degree $l$, denoted by ${P_{l}}$. Thus, to streamline notation, the following shorthand is utilized:
\begin{subequations}
	\begin{align}
	&\Psi _i^{l,0}(r) = \Psi _i^l(r),\label{eq42a}\\
	&\Psi _i^{l,0;\wp } = \Psi _i^{l;\wp }\label{eq42b},
	\end{align}
\end{subequations}
where $\Psi$ applies to any of the MEE field quantities.

The following spectral analysis (\cite{f50}) is valid for both cases of polymer matrix (dielectric) and metal matrix (plasmonics).  Assuming time harmonic incident P wave propagating through such media, then in view of the definitions given in Eqs.~\eqref{eq1} and \eqref{eq11} the incident displacement, ${\mathbf{\bar u}^{\mathcal{I}(2)}}(\mathbf{x},t)$ within the matrix of interest may be presented as: 
\begin{align}\label{eq800}
{\mathbf{\bar u}^{\mathcal{I}(2)}}(\mathbf{x},t) = \Lambda {e^{\iota ({K_p}z - \omega t)}}{\mathbf{e}_z} & = \Lambda {e^{\iota {K_p}z}}{e^{ - \iota \omega t}}{\mathbf{e}_z}\notag \\
&= {\mathbf{u}^{\mathcal{I}(2)}}(\mathbf{x}){e^{ - \iota \omega t}},
\end{align}
where the amplitude of the incident displacement, ${\mathbf{u}^{I(2)}}(\mathbf{x})$ can be expanded in terms of the VSHs as follows:
\begin{equation}\label{eq4004}
{\mathbf{u}^{\mathcal{I}(2)}}(\mathbf{x})= \sum\limits_{l = 0}^\infty  {[u_1^{l;\mathcal{I}(2)}(r)\mathbf{V}_1^l(\mathit{\Omega} ) + u_2^{l;\mathcal{I}(2)}(r)\mathbf{V}_2^l(\mathit{\Omega} )]}.
\end{equation}
It turns out that the coefficients in the above series become identically equal to become:
\begin{align}
&u_1^{l;\mathcal{I}(2)} = \frac{{ - (2l + 1){\iota ^{l + 1}}}}{{{K_P}}}\frac{{\partial {j_l}({K_P}r)}}{{\partial r}},\label{eq4005}\\
&u_2^{l;\mathcal{I}(2)}(r) =  - (2l + 1){\iota ^{l + 1}}\frac{{{j_l}({K_P}r)}}{{{K_P}r}}\label{4006},
\end{align}
in which ${{j_l}}$ is the spherical Bessel function of the first kind of of order $l$.

With the aids of the first part of Eq.~\eqref{eq2014}, Eq.~\eqref{eq998}, and writing ${\nabla _{{\mathbb{R}^3}}}$ with respect to the spherical coordinates, the following spectral relations between the generalized coordinates of strain and displacements in the region ${\Gamma _1} \cup {\Gamma _2}$ are obtained:
\begin{equation}\label{eq2}
\begin{split}
&\epsilon _1^{l}(r) = \frac{{du_1^{l}(r)}}{{dr}},\\
&\epsilon _2^{l}(r) = \frac{{u_1^{l}(r)}}{r} - \frac{1}{2}l(l + 1)\frac{{u_2^{l,m}(r)}}{r},\\
&\epsilon _3^{l}(r) = \frac{1}{2}\frac{{u_1^{l}(r)}}{r} + \frac{1}{2}r\frac{d}{{dr}}\left(\frac{{u_2^{l}(r)}}{r}\right),\\
&\epsilon _4^{l}(r) = \frac{{u_2^{l}(r)}}{r},\\
&\epsilon _5^l(r) = \frac{1}{2}r\frac{d}{{dr}}\bigg(\frac{{u_3^l(r)}}{r}\bigg),\\
&\epsilon _6^l(r) =  - \frac{{u_3^l(r)}}{r}.
\end{split}
\end{equation}
Likewise, Eqs.~\eqref{eq419}, \eqref{eq2015}, and \eqref{eq2018} lead to the following spectral relations between the generalized coordinates in the spectral expansions of the strain and displacement fields pertinent to the surface / interface boundaries:
\begin{equation}\label{3}
\begin{split}
&\epsilon _1^{l;\wp}(R) = 0, \hfill \\
&\epsilon _2^{l;\wp}(R) = \frac{{u_1^{l;\wp}(R)}}{R} - \frac{1}{2}l(l + 1)\frac{{u_1^{l;\wp}(R)}}{R}, \hfill \\
&\epsilon _3^{l;\wp}(R) = 0, \hfill \\
&\epsilon _4^{l;\wp}(R) = \frac{{u_2^{l;\wp}(R)}}{R}, \hfill \\
&\epsilon _5^{l;\wp}(R) = 0, \hfill \\
&\epsilon _6^{l;\wp}(R) =  - \frac{{u_3^{l;\wp}(R)}}{R}. \hfill
\end{split}
\end{equation}
Here, $R = {R_1}$ if $\wp$ represents the free surface and  $R ={R_2}$ if $\wp$ represents the interface.

The spectral constitutive relations for the bulk of the nanospherical MEE shell which is assumed to have a spherically isotropic properties, can be derived directly from Eqs.~\eqref{eq998}, \eqref{eq2016}, \eqref{eq2017}, and \eqref{eq2014}:
\begin{align}
&\sigma _1^{l}(r) = {C_{33}}\epsilon _1^{l}(r) + 2{C_{13}}\epsilon _2^{l}(r) - {e_{33}}E_1^{l}(r) - {\gamma_{33}}H_1^{l}(r),\notag\\
&\sigma _2^{l}(r) = {C_{13}}\epsilon _1^{l}(r) + \big({C_{11}} + {C_{12}}\big)\epsilon _2^{l}(r) - {e_{31}}E_1^{l}(r)- {\gamma_{31}}H_1^{l}(r),\notag\\
&\sigma _3^{l}(r) = 2{C_{44}}\epsilon _3^{l}(r) - {e_{15}}\big(E_2^{l}(r)+E_3^{l}(r)\big)- {\gamma_{15}}\big(H_2^{l}(r)+H_3^{l}(r)\big),\notag\\
&\sigma _4^{l}(r) = \big ({C_{11}} - {C_{12}}\big )\epsilon _4^{l}(r),\notag\\
&\sigma _5^l(r) = 2{C_{44}}\epsilon _5^l(r),\notag\\
&\sigma _6^l(r) = \big({C_{11}} - {C_{12}}\big)\epsilon _6^l(r),\label{eq4000}\\
&D_1^{l,m}(r) = {e_{33}}\epsilon _1^{l,m}(r) + 2{e_{31}}\epsilon _2^{l}(r) + {{\kappa _{33}}}E_1^{l}(r) + {{\mathscr{G} _{33}}}H_1^{l}(r),\notag\\
&D_2^{l}(r) = 2{e_{15}}\epsilon _3^{l}(r) + {{\kappa _{11}}}E_2^{l}(r) + {{\mathscr{G} _{11}}}H_2^{l}(r),\notag\\
&D_3^{l}(r) = 2{e_{15}}\epsilon _5^{l}(r)+{{\kappa _{11}}}E_3^{l}(r)+{{\mathscr{G} _{11}}}H_3^{l}(r),\notag\\
&B_1^{l,m}(r) = {\gamma_{33}}\epsilon _1^{l,m}(r) + 2{\gamma_{31}}\epsilon _2^{l}(r) + {{\mu _{33}}}H_1^{l}(r) + {{\mathscr{G} _{33}}}E_1^{l}(r),\notag\\
&B_2^{l}(r) = 2{\gamma_{15}}\epsilon _3^{l}(r) + {{\mu _{11}}}H_2^{l}(r) + {{\mathscr{G} _{11}}}E_2^{l}(r),\notag\\
&B_3^{l}(r) = 2{\gamma_{15}}\epsilon _5^{l}(r)+{{\mu _{11}}}H_3^{l}(r)+{{\mathscr{G} _{11}}}E_3^{l}(r).\notag
\end{align}
The distinction between the above constitutive relations and those employed recently in the work of \cite{f50} lies in the fact that the current work is concerned with a nanospherical MEE particle as opposed to the case of a nanospherical piezoelectric particle considered in \cite{f50}.

Likewise, using Eqs.~\eqref{eq997}, \eqref{eq2016}, \eqref{eq2017}, and \eqref{eq2014}, the spectral constitutive relations of the isotropic dielectric/metal matrix in the spherical coordinate system are given by
\begin{equation}\label{eq4001}
\begin{split}
&\sigma _1^l(r) = (\lambda  + 2\mu )\epsilon _1^l(r) + 2\lambda \epsilon _2^l(r),\\
&\sigma _2^l(r) = \lambda \epsilon _1^l(r) + 2(\lambda  + \mu )\epsilon _2^l(r),\\
&\sigma _3^l(r) = 2\mu \epsilon _3^l(r),\\
&\sigma _4^l(r) = 2\mu \epsilon _4^l(r),\\
&\sigma _5^l(r) = 2\mu \epsilon _5^l(r),\\
&\sigma _6^l(r) = 2\mu \epsilon _6^l(r),\\
&D_1^l(r) = {{\kappa _{m}}}E_1^l(r),\\
&D_2^l(r) = {{\kappa _{m}}}E_2^l(r),\\
&D_3^l(r) = {{\kappa _{m}}}E_3^l(r),\\
&B_1^l(r) = {\mu _m}H_1^l(r),\\
&B_2^l(r) = {\mu _m}H_2^l(r),\\
&B_3^l(r) = {\mu _m}H_3^l(r).
\end{split}
\end{equation}

It should be emphasized that in the present work, the surface/interface boundaries are assumed to have spherically isotropic properties.  Moreover, in addition to the surface/interface residual stresses, the surface / interface polarization and magnetization are also taken into account. To this end, by the utilization of Eqs.~\eqref{eq370}-\eqref{eq372}, \eqref{eq2015}, and \eqref{eq2018} the following spectral constitutive relations pertinent to the mentioned surface/interface boundaries are obtained:
\begin{align}
&\sigma _2^{l;\wp} = (C_{11}^\wp + C_{12}^\wp)\epsilon _2^{l;\wp},\label{eq34}\\
&\sigma _4^{l;\wp} = (C_{11}^\wp - C_{12}^\wp)\epsilon _4^{l;\wp},\label{eq340}\\
&\sigma _6^{l;\wp} = (C_{11}^\wp - C_{12}^\wp)\epsilon _6^{l;\wp},\label{eq341}\\
&P_2^{l;\wp} = \kappa_{11}^\wp E_2^{l;\wp}, \label{eq35}\\
&P_3^{l;\wp} = \kappa_{11}^\wp E_3^{l;\wp}.\label{eq36}\\
&M_2^{l;\wp} = \mu_{11}^\wp H_2^{l;\wp}, \label{eq37}\\
&M_3^{l;\wp} = \mu_{11}^\wp H_3^{l;\wp}.\label{eq38}
\end{align}
as discussed in Section~\ref{Sec3} $\sigma _1^{l;\wp}=\sigma _3^{l;\wp}=\sigma _5^{l;\wp} =P_1^{l;\wp}=M_1^{l;\wp}=0$

By expanding the MEE fields in terms of the VSHs and TSHs, the spectral representation of the fully coupled elastodynamics and Maxwell's equations \eqref{eq5000}, \eqref{eq1000} - \eqref{eq201} can be written as
\begin{align}
&\frac{{d\sigma _1^{l}(r)}}{{dr}} + 2\frac{{\sigma _1^{l}(r)}}{r} - 2\frac{{\sigma _2^{l}(r)}}{r} - l(l + 1)\frac{{\sigma _3^{l}(r)}}{r}=- \rho {\omega ^2}u_1^l(r),\label{eq790}\\
&\frac{{d\sigma _3^{l}(r)}}{{dr}} + \frac{{\sigma _2^{l}(r)}}{r} + 3\frac{{\sigma _3^{l}(r)}}{r}-\frac{1}{2}(l - 1)(l + 2)\frac{{\sigma _4^{l}(r)}}{r}= - \rho {\omega ^2}u_2^l(r),\label{eq80}\\
&\frac{{d\sigma _5^{l}(r)}}{{dr}} + 3\frac{{\sigma _5^{l}(r)}}{r} + \frac{1}{2}(l - 1)(l + 2)\frac{{\sigma _6^{l}(r)}}{r}+ \rho b_3^{l}(r) =- \rho {\omega ^2}u_3^l(r),\label{eq81}\\
&l(l + 1)\frac{{H_3^l(r)}}{r} = \iota \omega D_1^l(r),\refstepcounter{equation}\subeqn \label{eq4010}\\
&\frac{{dH_3^l(r)}}{{dr}} + \frac{{H_3^l(r)}}{r} = \iota \omega D_2^l(r),\subeqn \label{eq4011}\\
&\frac{{H_1^l(r)}}{r} - \frac{{dH_2^l(r)}}{{dr}} - \frac{{H_2^l(r)}}{r} = \iota \omega D_3^l(r),\subeqn \label{eq4012}\\
&l(l + 1)\frac{{E_3^l(r)}}{r} =  - \iota \omega B_1^l(r),\subeqn \label{eq4013}\\
&\frac{{dE_3^l(r)}}{{dr}} + \frac{{E_3^l(r)}}{r} =  - \iota \omega B_2^l(r),\subeqn \label{eq4014}\\
&\frac{{dE_2^l(r)}}{{dr}} + \frac{{E_2^l(r)}}{r} - \frac{{E_1^l(r)}}{r} = \iota \omega B_3^l(r).\subeqn \label{eq4015}
\end{align}
It is noteworthy to mention that, in contrast to the case of nanospherical piezoelectric particle (\cite{f50}) where $\{{\sigma _1}, {\sigma _2}, {\sigma _3}, {D_1}, {D_2}, {D_3}\}$ and $\{{B_1}, {B_2}, {B_3}\}$ within the particle depend, respectively, on $\{\boldsymbol{\epsilon},\mathbf{E}\}$ and $\{\mathbf{H}\}$, for the case of MEE nanospherical particle considered in the current work the multiphysics coupled field quantities $\{{\sigma _1}, {\sigma _2}, {\sigma _3}, {D_1}, {D_2}, {D_3},{B_1}, {B_2}, {B_3}\}$ within the particle depend on $\{\boldsymbol{\epsilon}, \mathbf{E}, \mathbf{H}\}$ concurrently.  The noted dependencies can be clearly observed from the constitutive relations provided by Eqs.~\eqref{eq4001}.  Moreover, contrary to the case considered by (\cite{f50}) where the surface / interface magnetization was absent, in the current case of MEE particle the surface / interface magnetization is present and has been accounted for properly.  Such a phenomenon together with the surface/interface polarization and residual stresses enter our mathematical formulations through the non-standard surface/interface conditions given by Eqs.~\eqref{eq2002}, \eqref{eq20000}, \eqref{eq422}, and \eqref{eq423} which will conveniently be employed using EIM method.
Another notable feature of the current work is the mathematical modeling of plasmonic matrix which surrounds the nanospherical MEE particle.  The optical properties of the metal matrix can be explained by plasmonic model through which the dielectric function depends on angular frequency.

The above coupled spectral ODEs must be solved with respect to some nonstandard spectral boundary conditions arisen from the consideration of surface/interface polarization and magnetization in addition to the surface/ interface residual stresses. The remainder of this section is devoted to the derivation and discussion of these nonstandard boundary conditions. By expanding the MEE fields appearing in Eqs.~\eqref{eq2002}, \eqref{eq20000}, \eqref{eq422}, and \eqref{eq423} in terms of spherical harmonics as outlined in Eqs.~\eqref{eq2016}-\eqref{eq2018}, the corresponding nonstandard spectral surface/interface boundary conditions are obtained as follows:
\begin{align}
&[{u_1}] = [{u_2}] = [{u_3}] = 0 \qquad \quad \text{on}\hspace{0.5em}r = {R_2},\label{eq43}\\
&[\sigma _1^l(R)] - 2\frac{{\sigma _2^{l;\wp}(R)}}{R} =  - {\rho ^\wp}{\omega ^2}u_1^{l}(R) \qquad \quad \label{eq44}\\&\qquad \quad\qquad \quad\text{on}\hspace{0.5em}r = {R_1} \hspace{0.5em} \text{and}\hspace{0.5em} r = {R_2},\notag\\
&[\sigma _3^l(R)] + \frac{1}{R}\bigg (\sigma _2^{l;\wp}(R) - \frac{1}{2}(l - 1)(l + 2)\sigma _4^{l;\wp}(R)\bigg )\label{eq45}\\
&\qquad\qquad\qquad\qquad\qquad\qquad\qquad=  - {\rho ^\wp}{\omega ^2}u_2^{l}(R)\notag\\&\qquad\qquad\qquad \text{on}\hspace{0.5em}r = {R_1} \hspace{0.5em} \text{and}\hspace{0.5em} r = {R_2},\notag\\
&[\sigma _5^l(R)] + \frac{1}{{2R}}(l - 1)(l + 2)\sigma _6^{l;\wp}(R) =  - {\rho ^\wp}{\omega ^2}u_3^{l}(R)\label{eq450}\\&\qquad\qquad\qquad\text{on}\hspace{0.5em}r = {R_1} \hspace{0.5em} \text{and}\hspace{0.5em} r = {R_2},\notag\\
&E_2^{l + } =  - {Z_{11}}H_3^{l + } + {Z_{12}}H_3^{l - }\label{eq452}\\
&\qquad\qquad\qquad\text{on}\hspace{0.5em}r = {R_1} \hspace{0.5em} \text{and}\hspace{0.5em} r = {R_2},\notag\\
&E_2^{l - } =  - {Z_{12}}H_3^{l + } + {Z_{11}}H_3^{l - }\label{eq453}\\
& \qquad\qquad\qquad\text{on}\hspace{0.5em}r = {R_1} \hspace{0.5em} \text{and}\hspace{0.5em} r = {R_2},\notag\\
&E_3^{l + } = {Z_{11}}H_2^{l + } - {Z_{12}}H_2^{l - }\label{eq454}\\&\qquad\qquad\qquad\text{on}\hspace{0.5em}r = {R_1} \hspace{0.5em} \text{and}\hspace{0.5em} r = {R_2},\notag\\
&E_3^{l - } = {Z_{12}}H_2^{l + } - {Z_{11}}H_2^{l - }\label{eq430}\\
&\qquad\qquad\qquad\text{on}\hspace{0.5em}r = {R_1} \hspace{0.5em} \text{and}\hspace{0.5em} r = {R_2}.\notag
\end{align}
Eqs. \eqref{eq452} - \eqref{eq430} arise from consideration of the EIM method; in these equations the superscript "$+$" over a quantity indicates that the quantity belongs to $R_1^{l + }$ or $R_2^{l + }$, whereas "$-$" superscript refers to those of $R_1^{l + }$ or $R_2^{l + }$.
The remainder of this section provides the solution to the governing spectral BVP outlined by Eqs.~\eqref{eq790}-\eqref{eq4015} are subjected to the nonstandard spectral surface/interface boundary conditions, as defined in Eqs.~\eqref{eq43}-\eqref{eq430}. 
Considering that the electric and the magnetic fields are governed by Helmholtz equation, the uniqueness of solution inquires that their corresponding Sommerfeld radiation condition:
\begin{equation}
\mathop {\lim }\limits_{r \to \infty } (\frac{\partial }{{\partial r}} - \iota k)\psi (r) = 0
\end{equation}

As noted earlier, due to the rotational symmetry of the field quantities, the solutions of interest pertain to each of the harmonics with $m=0,\hspace{0.25em}l \geqslant 0$ (Eqs.~\eqref{eq42a} and \eqref{eq42b}), and thus the general solution may be obtained as a superposition of all solutions over $l$ while $m$ is kept equal to zero.

By the utilization of Eqs.~\eqref{eq2}, \eqref{eq4001}, \eqref{eq790}, and \eqref{eq80} the components of the scattered mechanical displacement field within the matrix are readily obtained:
\begin{align}
&{u_1^{l;\Im(2)}}(r) = {a_l}\frac{\partial }{{\partial r}}[{h_l}({K_P}r)] + {b_l}l(l + 1)\frac{{{h_l}({K_S}r)}}{r}\label{eq50},\\
&{u_2^{l;\Im(2)}}(r)= {a_l}\frac{{{h_l}({K_P}r)}}{r} + \frac{{{b_l}}}{r}\frac{\partial }{{\partial r}}[r{h_l}({K_S}r)]\label{eq51},
\end{align}
where ${a_l}$ and ${b_l}$, $l \geqslant 0$ are constants which will be determined by imposing the boundary conditions. It should be noted that the matrix is made of a material such that the scattered fields are decoupled; for this reason Eqs.~\eqref{eq50} and \eqref{eq51} for the matrix coincide with those obtained in \cite{m502} and \cite{f50}.

By employing equations \eqref{eq4001}, \eqref{eq4010} through \eqref{eq4015}, we obtain the following equations for the scattered electromagnetic fields:
\begin{align}
&\big(\frac{{{d^2}}}{{d{r^2}}} + \frac{2}{r}\frac{d}{{dr}} - \frac{{l(l + 1)}}{{{r^2}}} + {\mu _m}{g_m}{\omega ^2}\big)H_3^{l;\Im(2)}(r) = 0,\label{eq52}\\
&\big(\frac{{{d^2}}}{{d{r^2}}} + \frac{2}{r}\frac{d}{{dr}} - \frac{{l(l + 1)}}{{{r^2}}} + {\mu _m}{g_m}{\omega ^2}\big)E_3^{l;\Im(2)}(r) = 0.\label{eq53}
\end{align}
Considering Sommerfeld's radiation condition, the solution to the aforementioned equation is as follows:
\begin{align}
&H_3^{l;\Im(2)}(r) = {\chi _l}{h_l}({K_m}r),\label{eq54}\\
&E_3^{l;\Im(2)}(r) = {\vartheta _l}{h_l}({K_m}r)\label{eq55}.
\end{align}
Here, ${K_m} = \frac{\omega }{{{C_m}}}$  represents the electromagnetic wave number of the matrix, where ${C_m} = \frac{1}{{\sqrt {{\mu _m}{{\kappa}_m}} }}$ denotes the speed of light in the matrix. Utilizing Eqs.~\eqref{eq4001}, \eqref{eq4010}, \eqref{eq4011}, \eqref{eq4013}, and \eqref{eq4014}, the spectral coefficients corresponding to the electric and the magnetic fields within the matrix are obtained:
\begin{align}
&E_1^{l;\Im(2)}(r) =  - \frac{{\iota {\chi _l}l(l + 1)}}{{r\omega {\kappa_m}}}{h_l}({K_m}r),\label{eq56}\\
&E_2^{l;\Im(2)}(r) = - \frac{{\iota {\chi _l}}}{{\omega {\kappa_m}}}\big(\frac{{\partial {h_l}({K_m}r)}}{{\partial r}} + \frac{{{h_l}({K_m}r)}}{r}\big),\label{eq57}\\
&H_1^{l;\Im(2)}(r) = \frac{{\iota {\vartheta _l}l(l + 1)}}{{r\omega {\mu _m}}}{h_l}({K_m}r),\label{eq58}\\
&H_2^{l;\Im(2)}(r) = \frac{{\iota {\vartheta _l}}}{{\omega {\mu_m}}}\big(\frac{{\partial {h_l}({K_m}r)}}{{\partial r}} + \frac{{{h_l}({K_m}r)}}{r}\big).\label{eq59}
\end{align}

In the solutions corresponding to the refracted electric and magnetic fields inside the MEE particle, the cases associated with $l=0$ and $l \geqslant 1$ will be treated separately.  To this end, first the solutions for the case of $l=0$ will be provided, and then the solutions for all the other terms corresponding to $l \geqslant 1$ will be treated in a unified manner.  For the case of $l = 0$, ${Y^{0,0}}(\mathit{\Omega} ) = 1$ and thus Eqs.~\eqref{eq7-1}, \eqref{eq7-2}, and \eqref{eq7-3} yield: 
\begin{equation}
\mathbf{V}_1^0(\Omega ) = 1{\mathbf{e}_r},\quad
\mathbf{V}_2^0(\Omega ) = 0,\quad
\mathbf{V}_3^0(\Omega ) = 0.
\end{equation}
Furthermore, since that the $\mathscr{Q}$ curl of any vector field in the form  $f(r){\mathbf{e}_r}$ equals zero, then the spectral strain-displacement relations \eqref{eq2}, and Eqs.~\eqref{eq4010} and \eqref{eq4013} yield:
\begin{align}
&\epsilon _1^{0;\Re(1)}(r) = \frac{{du_1^{0;\Re(1)}(r)}}{{dr}},\label{eq151}\\
&\epsilon _2^{0;\Re(1)}(r) = \frac{{u_1^{0;\Re(1)}(r)}}{r},\label{eq152}\\
&D_1^{0;\Re(1)}(r) = 0,\label{eq153}\\
&B_1^{0;\Re(1)}(r) = 0.\label{eq154}
\end{align}
Now, using Eqs.~\eqref{eq153}, and \eqref{eq154}, we obtain 
\begin{align}
&E_1^{0;\Re(1)}(r) = \frac{{2({e_{31}}{\mu _{33}} - {\gamma _{31}}{d_{33}})u_1^{0;\Re(1)}(r)}}{{r(d_{33}^2 - {g_{33}}{\mu _{33}})}} + \frac{{({e_{33}}{\mu _{33}} - {\gamma _{33}}{d_{33}})\frac{{du_1^{0;\Re(1)}}}{{dr}}}}{{d_{33}^2 - {g_{33}}{\mu _{33}}}}\label{eq155}\\
&H_1^{0;\Re(1)}(r) = \frac{{2({\gamma _{31}}{g_{33}} - {e_{31}}{d_{33}})u_1^{0;\Re(1)}(r)}}{{r(d_{33}^2 - {g_{33}}{\mu _{33}})}}+ \frac{{({\gamma _{33}}{g_{33}} - {e_{33}}{d_{33}})\frac{{du_1^{0;\Re(1)}}}{{dr}}}}{{d_{33}^2 - {g_{33}}{\mu _{33}}}}\label{eq156}
\end{align}
Also, considering Eq.~\eqref{eq4000} the constitutive equations will take the form
\begin{align}
&\sigma _1^{0;\Re(1)}(r) = {C_{33}}\epsilon _1^{0;\Re(1)}(r) + 2{C_{13}}\epsilon _2^{0;\Re(1)}(r) - {e_{33}}E_1^{0;\Re(1)} - {\gamma _{33}}H_1^{0;\Re(1)}\label{eq157}\\
&\sigma _2^{0;\Re(1)}(r) = {C_{13}}\epsilon _1^{0;\Re(1)}(r) + ({C_{11}} + {C_{12}})\epsilon _2^{0;\Re(1)}(r) - {e_{31}}E_1^{0;\Re(1)} - {\gamma _{31}}H_1^{0;\Re(1)}\label{eq158}
\end{align}
Combining Eqs.~\eqref{eq155}-\eqref{eq158} and the mechanical equilibrium equation\eqref{eq790} lead to following Helmholtz equation:
\begin{equation}\label{eq160}
{r^2}\frac{{{d^2}u_1^{0;\Re(1)}}}{{d{r^2}}} + 2r\frac{{du_1^{0;\Re(1)}}}{{dr}} - {\tilde{\mathcal{A}}}u_1^{0;\Re(1)} =  - {{\tilde{\mathcal{B}}}^2}{r^2}u_1^{0;\Re(1)}
\end{equation}
in which:
\begin{align}
{\tilde{\mathcal{A}}} = \frac{\Im }{\aleph },\\
{{\tilde{\mathcal{B}}}^2} = \frac{\Theta }{\aleph },
\end{align}
where:
\begin{align}
&\Im  = 2\bigg(\big(d_{33}^2 - {g_{33}}{\mu _{33}}\big)\big({C_{11}} + {C_{12}} - {C_{13}}\big)\notag\\
&\qquad + {d_{33}}\big({e_{31}}(4{\gamma _{31}} - {\gamma _{33}}) - {e_{33}}{\gamma _{31}}\big)\notag\\
&\qquad - {g_{33}}{\gamma _{31}}\big(2{\gamma _{31}} - {\gamma _{33}}\big) - {e_{31}}{\mu _{33}}\big(2{e_{31}} - {e_{33}}\big)\bigg),
\end{align}
\begin{align}
&\aleph  = {d_{33}}\big({C_{33}}{d_{33}} + 2{e_{33}}{\gamma _{33}}\big) - {g_{33}}\big(\gamma _{33}^2 + {C_{33}}{\mu _{33}}\big)\notag\\
&\qquad - {\mu _{33}}e_{33}^2,
\end{align}
\begin{equation}
\Theta  = \rho {\omega ^2}\big(d_{33}^2 - {g_{33}}{\mu _{33}}\big).
\end{equation}
The solution of Eq. \eqref{eq160} is:
\begin{equation}\label{eq1290}
{u_1}^{0;\Re(1)}(r) = {\alpha_5}{j_\varsigma }({{\tilde{\mathcal{B}}}}r)+{\alpha_6}{y_\varsigma }({{\tilde{\mathcal{B}}}}r),
\end{equation}
where ${j_\varsigma}({{\tilde{\mathcal{B}}}}r)$ and ${y_\varsigma }({{\tilde{\mathcal{B}}}}r)$ are, respectively, the spherical Bessel functions of the first and second kind of order $\varsigma$, with
\begin{equation}
\varsigma = \frac{{\sqrt {1 + 4{{\tilde{\mathcal{A}}}}}  - 1}}{2}.
\end{equation}
Hence, by utilizing Eqs. \eqref{eq155}, \eqref{eq156}, and \eqref{eq1290} the zeroth terms of the spectral components of the electric and magnetic fields $E_1^{0;\Re(2)}(r)$ and $H_1^{0;\Re(2)}(r)$ inside the MEE particle can be obtained.

Having obtained the zeroth term ($l=0$) in the solutions of the refracted magneto-electro-mechanical fields inside the MEE particle, we now turn attention to obtaining the terms of the solutions pertinent to $l \geqslant 1$.  To this end, for each $l \geqslant 1$ Eqs.~\eqref{eq2}, \eqref{eq4000}, \eqref{eq790}-\eqref{eq81}, and \eqref{eq4010}-\eqref{eq4015} are combined to get a system of linear ODEs.  For a typical term, $\mathscr{L}$ the system of ODEs may be represented in the following compact form:
\begin{align}
&{\mathbf{Q_0}}{r^2}\frac{{{d^2}{\mathbf{X}^{^{\mathscr{L};\Re (1)}}}(r)}}{{d{r^2}}} + ({\mathbf{Q_1}} + {\mathbf{Q_2}}r)r\frac{{{d}{\mathbf{X}^{^{\mathscr{L};\Re (1)}}}(r)}}{{d{r}}}\notag\\&\qquad\quad + ({\mathbf{Q_3}} + {\mathbf{Q_4}}r + {\mathbf{Q_5}}{r^2})\mathbf{X}^{^{\mathscr{L};\Re (1)}}(r) = 0,\label{eq110}
\end{align}
for the unknown MEE fields:  
\begin{equation}\label{eq111}
\mathbf{X}^{^{\mathscr{L};\Re (1)}}(r)= {\{\{ {u_1}(r),{u_2}(r),{E_3}(r),{H_3}(r)\} ^{\mathscr{L};\Re (1)}\}^T}.
\end{equation}
In Eqs.~\eqref{eq110},  ${\mathbf{Q}_{i}} = {q_{ijk}}{\mathbf{e}_j} \otimes {\mathbf{e}_k}$ where the coefficient ${q_{ijk}}$ contains a combination of the MEE parameters appearing in the MEE constitutive relations given in Eqs.~\eqref{eq4000}.
Eq.\eqref{eq110} can be readily converted into the following system of first order equations:
\begin{align}
& r\frac{{d{\mathbf{X}^{\mathscr{L};\Re (1)}}}}{r}-\mathbf{Z}^{\mathscr{L};\Re (1)} = 0,\label{eq112}\\
&{r}\frac{{d{\mathbf{Z}^{\mathscr{L};\Re (1)}}}}{dr} + \big(\mathbf{Q_0}^{ - 1}({\mathbf{Q_1}} + {\mathbf{Q_2}}r) - \mathbf{I}\big)r\mathbf{Z}^{\mathscr{L};\Re (1)} \notag\\&\qquad+ \mathbf{Q_0}^{ - 1}\big({\mathbf{Q_3}} + {\mathbf{Q_4}}r + {\mathbf{Q_5}}{r^2}\big)\mathbf{X}^{\mathscr{L};\Re (1)} = 0.\label{eq113}
\end{align}
The above system of equations can be recast into a more compact form as:
\begin{equation}\label{eq114}
r{\{ \frac{{d{\mathbf{X}^{\mathscr{L};\Re (1)}}}}{{dr}},\frac{{d{\mathbf{Z}^{\mathscr{L};\Re (1)}}}}{{dr}}\} ^T} = \mathbf{P}{\{ {\mathbf{X}^{\mathscr{L};\Re (1)}},{\mathbf{Z}^{\mathscr{L};\Re (1)}}\} ^T},
\end{equation}
where
\begin{equation}\label{eq115}
\mathbf{P}=
\begin{bmatrix}
0&\mathbf{I}\\
- \mathbf{Q_0}^{ - 1}({\mathbf{Q_3}} + {\mathbf{Q_4}}r + {\mathbf{Q_5}}{r^2})&\mathbf{I} - \mathbf{Q_0}^{ - 1}({\mathbf{Q_1}} + {\mathbf{Q_2}}r)
\end{bmatrix}.
\end{equation}
The solution to the system of the first order ODEs Eq.~\eqref{eq114} can be represented as the Frobenius series given below:
\begin{equation}\label{eq116}
{\{ \mathbf{X}^{\mathscr{L};\Re (1)},\mathbf{Z}^{\mathscr{L};\Re (1)}\} ^T} = \sum\limits_{i = 0}^\infty  {\{ \mathbf{F}_i^\mathbf{X},\mathbf{F}_i^\mathbf{Z}\}{r^{\xi  + i}}}
\end{equation}
where $\mathbf{F}_i^\mathbf{X}$ and $\mathbf{F}_i^\mathbf{Z}$ are the unknown coefficients in the Frobenious series for the vectors $\mathbf{X}$ and $\mathbf{Z}$, respectively. Subsequently, substitution of Eq.~\eqref{eq116} into Eq.~\eqref{eq114} yields:
\begin{align}
&\sum\limits_{i = 0}^\infty  {(\xi  + i){\mathbf{F}_i^\mathbf{X}}{r^{\xi  + i}}}  = {\mathbf{P_0}}\sum\limits_{i = 0}^\infty  {{\mathbf{F}_i^\mathbf{X}}{r^{\xi  + i}}}\notag\\&\qquad + {\mathbf{P_1}}\sum\limits_{i = 0}^\infty  {{\mathbf{F}_i^\mathbf{X}}{r^{\xi  + i + 1}}} + {\mathbf{P_2}}\sum\limits_{i = 0}^\infty  {{\mathbf{F}_i^\mathbf{X}}{r^{\xi  + i + 2}}}\label{eq118},
\end{align}
where
\begin{equation}\label{eq119}
\mathbf{{P_0}}=
\begin{bmatrix}
0&\mathbf{I}\\
- \mathbf{Q_0}^{ - 1}{\mathbf{Q_3}}&\mathbf{I} - \mathbf{Q_0}^{ - 1}{\mathbf{Q_1}}
\end{bmatrix},
\end{equation}
\begin{equation}\label{eq120}
\mathbf{P_1}=
\begin{bmatrix}
0&0\\
- \mathbf{Q_0}^{ - 1} {\mathbf{Q_4}}& - \mathbf{Q_0}^{ - 1} {\mathbf{Q_2}}
\end{bmatrix},
\end{equation}
\begin{equation}\label{eq121}
\mathbf{P_2}=
\begin{bmatrix}
0&0\\
- \mathbf{Q_0}^{ - 1}{\mathbf{Q_5}}&0
\end{bmatrix}.
\end{equation}
Consequently,
\begin{align}
&i = 0 \to \xi {\mathbf{F}_0^\mathbf{X}} = {\mathbf{P_0}}{\mathbf{F}_0^\mathbf{X}},\label{eq1210}\\
&i = 1 \to (\xi  + 1){\mathbf{F}_1^\mathbf{X}} = {\mathbf{P_0}}{\mathbf{F}_1^\mathbf{X}} + {\mathbf{P_1}}{\mathbf{F}_0^\mathbf{X}},\label{eq122}\\
&i \ge 2 \to (\xi  + i){\mathbf{F}_i^\mathbf{X}} = {\mathbf{P_0}}{\mathbf{F}_i^\mathbf{X}} + {\mathbf{P_1}}{\mathbf{F}_{i-1}^\mathbf{X}} + {\mathbf{P_2}}{\mathbf{F}_{i-2}^\mathbf{X}}.\label{eq1230}
\end{align}
As it appears from Eq.~\eqref{eq1210}, $\xi$ and $\mathbf{F}_0^\mathbf{X}$ are the eigenvalue and the eigenvector of $\mathbf{P_0}$. From Eq.~\eqref{eq1230}, it follows that for $i \ge 2 $:
\begin{equation}
{\mathbf{F}_i} = {\big((\xi  + i)\mathbf{I} - {\mathbf{P}_0}\big)^{ - 1}}{\mathbf{P}_1}{\mathbf{F}_{i - 2}}.
\end{equation}
By applying this recursive formula, we can obtain the general solution to Eq.~\eqref{eq114}. The determination of the unknown coefficients $\mathbf{F}_i^\mathbf{X}$ will be accomplished by imposing the boundary conditions given by Eqs.~\eqref{eq2002}, \eqref{eq20000}, \eqref{eq422}, and \eqref{eq423}.

Assuming that the core region of the MEE nanospherical particle surrounding the origin, ${\Gamma^{(0)}}$ is vacuumed, the refracted electromagnetic fields in this region may be written as:
\begin{align}
&\big(\frac{{{d^2}}}{{d{r^2}}} + \frac{2}{r}\frac{d}{{dr}} - \frac{{l(l + 1)}}{{{r^2}}} + {\mu _0}{\kappa_0}{\omega ^2}\big)H_3^{l;\Re(0)}(r) = 0,\label{eq520}\\
&\big(\frac{{{d^2}}}{{d{r^2}}} + \frac{2}{r}\frac{d}{{dr}} - \frac{{l(l + 1)}}{{{r^2}}} + {\mu _0}{\kappa_0}{\omega ^2}\big)E_3^{l;\Re(0)}(r) = 0,\label{eq521}
\end{align}
for $i \ge 0 $. The solution of the above equations in  ${\Gamma^{(0)}}$ are given in terms of spherical Bessel functions as below:
\begin{align}
&H_3^{l;\Re(0)}(r) = {\chi _l}{j_l}({K_0}r),\label{eq540}\\
&E_3^{l;\Re(0)}(r) = {\vartheta _l}{j_l}({K_0}r).\label{eq541}
\end{align}
Here we used the fact that the solution must be bounded at $r=0$.
\section{Poynting Vector, Scattering Cross Section and Asymptotic Relations}
The analysis of this section is applicable to both metal and polymer matrix. Now, we will derive a simple expression for the power of the scattered EM waves in the matrix. The total time-averaged EM radiated power, $\left\langle \mathscr{P} \right\rangle$ crossing a closed surface $\Sigma$ surrounding the origin is given by (\cite{f1})
\begin{equation}\label{eq143}
\left\langle \mathscr{P} \right\rangle  = \int_\Sigma  {\frac{1}{2}} {\mathop{\rm Re}\nolimits} \big[\mathbf{E} \times {\mathbf{H}^*}\big]\cdot\mathbf{n}\hspace{0.1em}dA,
\end{equation}
where $\mathbf{H}^*$ is the complex conjugate of $\mathbf{H}$, and $\textrm{Re} [.]$ denotes the real part of $[.]$.  Moreover, $\mathbf{n}$ is the unit outward normal to the bounding surface $\Sigma$ with the element of area $dA$. Now, considering that the spherical Hankel functions satisfy the asymptotic relation
\begin{equation}\label{eq144}
{h_l}({K_m}r) \approx {( - \iota )^{l + 1}}\frac{{{e^{\iota {K_m}r}}}}{{{K_m}r}}\qquad\text{as}\qquad r \to \infty,
\end{equation}
it follows from Eqs.~\eqref{eq54}-\eqref{eq59} that as $r \to \infty$,
\begin{align}
&E_2^l(r) \approx {\chi _l}{( - \iota )^{l + 1}}\frac{{{e^{\iota {K_m}r}}}}{{r\omega {g_m}}},\label{eq145}\\
&E_3^l(r) \approx {\vartheta _l}{( - \iota )^{l + 1}}\frac{{{e^{\iota {K_m}r}}}}{{{K_m}r}},\label{eq146}\\
&H_2^l(r) \approx  - {\vartheta _l}{( - \iota )^{l + 1}}\frac{{{e^{\iota {K_m}r}}}}{{r\omega {\mu _m}}},\label{eq147}\\
&H_3^l(r) \approx {\chi _l}{( - \iota )^{l + 1}}\frac{{{e^{\iota {K_m}r}}}}{{{K_m}r}}.\label{eq148}
\end{align}
Making use of the norms:
\begin{equation}
{\left\| {\mathbf{V}_2^{l,0}} \right\|} = {\left\| {\mathbf{V}_3^{l,0}} \right\|}= \sqrt{4\pi \frac{{l(l + 1)}}{{2l + 1}}}.
\end{equation}
and Eqs.~\eqref{eq145} - \eqref{eq148}, the following expression for $\left\langle \mathscr{P} \right\rangle$ is obtained:
\begin{equation}\label{eq1150}
\left\langle \mathscr{P} \right\rangle  = \frac{{2\pi }}{{\sqrt {{g_m}{\mu _m}} {\omega ^2}}}\sum\limits_{l = 1}^\infty  {\frac{{l(l + 1)}}{{2l + 1}}} (\frac{{{{\left| {{\chi _l}} \right|}^2}}}{{{g_m}}} + \frac{{{{\left| {{\vartheta _l}} \right|}^2}}}{{{\mu _m}}}).
\end{equation}
The EM radiated power is conveniently normalized as follows
\begin{align}
&\left\langle {\mathscr{\hat P}} \right\rangle  = \frac{{\left\langle \mathscr{P} \right\rangle }}{{\left\langle {{\mathbf{I}_0}} \right\rangle  }},\label{eq196}\\
&\left\langle {{\mathbf{I}_0}} \right\rangle  = \frac{1}{2} \sqrt {{\rho _m}(\lambda  + 2\mu )} \pi {\omega ^2}{R^2}{\left| {{\Lambda}} \right|^2},\label{eq197}
\end{align}
in which $\left\langle {\mathscr{\hat P}} \right\rangle$ is also called antenna efficiency.
In the above equation $\left\langle {{\mathbf{I}_0}} \right\rangle$ represents the time-averaged energy flux of the incident P-wave. This flux passes through the cross-sectional area of the nano-spherical particle, projected onto the xy-plane (the plane perpendicular to the propagation direction).
\section{Results and discussion}
In this section, we demonstrate the generality and robustness of our approach by solving examples of varying physical complexity. Unlike methods relying on the electroquasistatic approximation, our approach incorporates not only the fully coupled dynamic Maxwell and elastodynamic equations but also accounts for the surface effects properly and thus provides accurate predictions of electromagnetic wave propagation. It is particularly effective for nanoscale MEE problems, where surface and interface electromagnetization effects are adequately managed using surface elasticity theory combined with the EIM method.Since the size of the MEE shell is in the range of only a few nanometers, then we focus on incident P-waves at THz frequencies.

This section states and discusses three illustrative examples. The first example which revisits the case of a piezoelectric shell embedded in a polymer matrix subjected to P-waves considered recently by (\cite{f50}) serves as a verification of the current formulations. The remaining examples introduce novel applications, focusing on technological advancements in antennas and filters. Specifically, they explore MEE fields in nano-sized antennas subjected to acoustic waves, analyzing how surface and interface electromechanical properties together with the volume fractions of the nano-shell’s constituents affect antenna efficiency, bandwidth, and resonance frequency.

We introduce the following interface characteristic lengths: ${{\hat m}_1} = \frac{{C_{11}^\wp }}{{{C_{11}}}}$, ${{\hat m}_2} = \frac{{C_{12}^\wp }}{{{C_{12}}}}$, ${{\hat m}_3} = \frac{{{\rho ^\wp }}}{\rho }$, ${{\hat m}_4} = \frac{{\kappa _{11}^\wp }}{{{\kappa _{11}}}}$.  If one or more of these length scales is comparable to the size of the MEE shell or the wavelength of the incident wave, the surface and interface effects become significant (\cite{f58}). Conversely, if $R_1$ and $R_2$ are large compared to these characteristic lengths, then for large wavelengths, the nanoscopic features of the MEE shell such as the surface and the interface effects become negligibly small. Consequently, the analysis will be reduced to that given based on classical theory.
\subsection{Example 1: PZT-4 shell embedded in epoxy matrix}\label{sec10}
For the sake of verification of the current theory and formulations, this example examines the special case of a nano-sized piezoelectric shell respectively, with the inner and outer radii, $R_1=0.5$ nm and $R_2=1$ nm embedded in a polymer matrix subjected to incident P-waves as considered by \cite{f50}.  In this set problem the piezoelectric shell modeled a nano-sized antenna.  They calculated the resulting variations in antenna efficiency in terms of the normalized incident wave number (${K_p}{R_2}$) for several different characteristic lengths ${{\hat m}_1} = {{\hat m}_2} = {{\hat m}_3},\hspace{0.25em}\text{and}\hspace{0.25em}{{\hat m}_4}$.  Clearly, in this special case as opposed to the more general case of the MEE shell treated in the current work, magnetization incurs neither within the bulk of the piezoelectric shell nor on its surface and interface domains. The electro-mechanical properties of the piezoelectric shell and the epoxy matrix are provided in Tables~\ref{tbl1} and \ref{tbl3}, respectively.

\begin{figure}[H]
	\begin{center}
		\includegraphics[width=12cm]{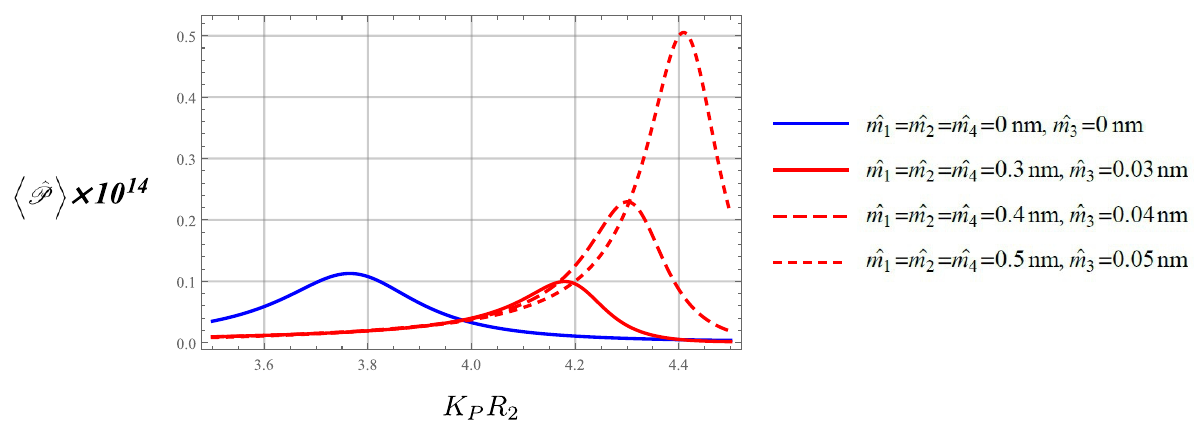}
	\end{center}
	\caption{Effects of characteristic lengths on variations in antenna efficiency for a nano-sized piezoelectric shell embedded in an epoxy matrix. The results are in exact agreement with those reported by \cite{f50}}
	\label{300000}
\end{figure}

The variations of the antenna efficiency versus normalized incident wavenumber which have been calculated for different characteristic lengths using the current formulations are displayed in Fig.~\ref{300000}.  These results are in exact agreement with those given in \cite{f50}.
\begin{table}[H]
	\centering
	\begin{subtable}[b]{0.3\linewidth}
		\begin{center}
			\begin{tabular}{c|c}
				\hline
				\hline
				Property
				&PZT-4\\
				\hline
				${C_{11}}(\text{GPa})$
				&139\\
				${C_{12}}$
				&77.8\\
				${C_{13}}$
				&74.3\\
				${C_{33}}$
				&115\\
				${C_{44}}$
				&25.6\\
				${e_{31}}(\text{C}/{\text{m}^2})$
				&-5.2\\
				${e_{33}}$
				&15.1\\
				${e_{15}}$
				&12.7\\
				${k_{11}}({10^{ - 10}}\text{F}/\text{m})$
				&64.64\\
				${k_{33}}$
				&56.22\\
				${\mu_{11}}({10^{ - 6}}\text{N}/{\text{A}^2})$
				&5\\
				${\mu_{33}}$
				&10\\
				$\rho(\text{Kg}/{\text{m}^3})$
				&7500\\
				\hline
				\hline
			\end{tabular}
		\end{center}
		\caption{}
		\label{tbl1}
	\end{subtable}
	\begin{subtable}[b]{0.6\linewidth}
		\begin{center}
			\begin{tabular}{c|c}
				\hline
				\hline
				Property
				&Epoxy\\
				\hline
				$\lambda(\text{GPa})$
				&4.916\\
				$\mu(\text{GPa})$
				&1.731\\
				${\kappa_m}({10^{ - 10}}\text{F}/\text{m})$
				&0.38\\
				${\mu_{m}}({10^{ - 6}}\text{N}/{\text{A}^2})$
				&2.51\\
				$\rho(\text{Kg}/{\text{m}^3})$
				&1202\\
				\hline
				\hline
			\end{tabular}
		\end{center}
		\caption{}
		\label{tbl3}
	\end{subtable}
	\caption{Electro-mechanical properties of (a) piezoelectric shell, and
		(b) epoxy matrix}
	\label{table}
\end{table}
\subsection{Example 2: Embedded MEE shell inside a metal matrix within plasmonic framework or inside a polymer matrix}
In this example, it is assumed that an embedded nano-sized spherical MEE shell composed of BaTiO3 (piezoelectric) and COFe2O4 (piezomagnetic) is subjected to an incident acoustic wave.  Two different types of host matrices made of (a) epoxy and (b) metal are considered.  In the latter case of the metallic matrix the interesting phenomenon of plasmonic are captured and its effects examined.  Moreover, in order to address the effects of the percent compositions of MEE shell on the antenna efficiency, two different compositions of $\mathrm{40\% BaTi{O_3}} - \mathrm{60\% Co{Fe_2}{O_4}}$ and $\mathrm{60\% BaTi{O_3}}-\mathrm{40\% Co{Fe_2}{O_4}}$ denoted, respectively, as  $\mathscr{C}_1$ and $\mathscr{C}_2$ will be examined in both cases of (a) and (b).  The properties for the bulk regions of the MEE shells pertinent to the compositions  $\mathscr{C}_1$ and $\mathscr{C}_2$ which are displayed in Table.~\ref{tbl13} are based on the data from \cite{f56} and \cite{f57}.  The electromechanical properties utilized in the cases of (a) polymer matrix and (b) aluminium matrix are given in Tables~\ref{tbl3} and \ref{tbl14}, respectively.
\begin{table*}
	\begin{center}
		\resizebox{14cm}{!}{
			\begin{tabular}{c|ccccccccccccccc}
				\hline
				\hline
				Material&&${C_{11}}$
				&${C_{12}}$&${C_{13}}$&${C_{33}}$&${C_{44}}$&&${e_{31}}$&${e_{33}}$&${e_{15}}$&&${q_{31}}$&${q_{33}}$&${q_{15}}$\\
				&&\multicolumn{5}{c}{($\text{GPa}$)}& &\multicolumn{3}{c}{($\text{C}/{\text{m}^2}$)}
				&&\multicolumn{3}{c}{(${10^{ - 9}}\text{F}/\text{m}$)}\\\cline{1-1} \cline{3-7}\cline{9-11}\cline{13-15}
				$\mathscr{C}_1$&&139
				&77.8&74.3&115&25.6&&-5.2&15.1&12.7&&6.464&5.622&2\\
				\cline{1-1}\cline{3-7}\cline{9-11}\cline{13-15}
				$\mathscr{C}_2$&&126&55&53&117&35.5&&-6.5&23.3&17&&15.1&13&2\\&&&&&&&&&&&&&\\\hline\\Material&&${e_{31}}$&&${e_{33}}$&&&${k_{11}}$&&${k_{33}}$&&&${k_{11}}$&&${k_{33}}$\\&&\multicolumn{3}{c}{($\text{GPa}$)}&&&\multicolumn{3}{c}{($\text{C}/{\text{m}^2}$)}&&&\multicolumn{3}{c}{($\text{C}/{\text{m}^2}$)}\\\cline{1-1}\cline{3-5}\cline{8-10}\cline{13-15}$\mathscr{C}_1$&&126&&53&&&-6.5&&17&&&15.1&&2\\\cline{1-1}\cline{3-5}\cline{8-10}\cline{13-15}$\mathscr{C}_2$&&126&&53&&&-6.5&&17&&&15.1&&2\\ \hline\hline
		\end{tabular}}
	\end{center}
	\caption{The electro-magneto-elastic properties of $\mathscr{C}_1$ and $\mathscr{C}_2$ compositions}
	\label{tbl13}
\end{table*}
\begin{table*}
	\begin{center}
		\begin{tabular}{c|c}
			\hline
			\hline
			Property
			&Aluminium\\
			\hline
			$\lambda(\text{GPa})$
			&60.493\\
			$\mu(\text{GPa})$
			&25.925\\
			${\omega_p}({10^{16}}\text{Hz})$
			&2.24\\
			${\mu_{m}}({10^{ - 6}}\text{N}/{\text{A}^2})$
			&1.256\\
			$\rho(\text{Kg}/{\text{m}^3})$
			&2700\\
			\hline
			\hline
		\end{tabular}
	\end{center}
	\caption{The electro-magneto-elastic properties of aluminium matrix}
	\label{tbl14}
\end{table*}
\subsubsection{Case (a) - MEE fields of a nano-sized antenna embedded in an epoxy matrix}\label{sec1a}
Consider a nano-sized MEE shell with an outer-to-inner radius ratio of 
$\frac{{{R_2}}}{{{R_1}}} = 2$ embedded in an epoxy matrix.
When acoustic waves interact with the shell, the presence of both the free surface and the interface of the ultra-thin shell with its surrounding matrix introduces interference from reflected waves within the shell thickness, significantly complicating the underlying physics. This complexity would be notably reduced if the antenna were constructed with a solid core. For this analysis, we assume the outer radius of the shell is $R_2 =1 \mathrm{nm}$. 

\begin{figure}[H]
	\begin{center}
		\includegraphics[width=12cm]{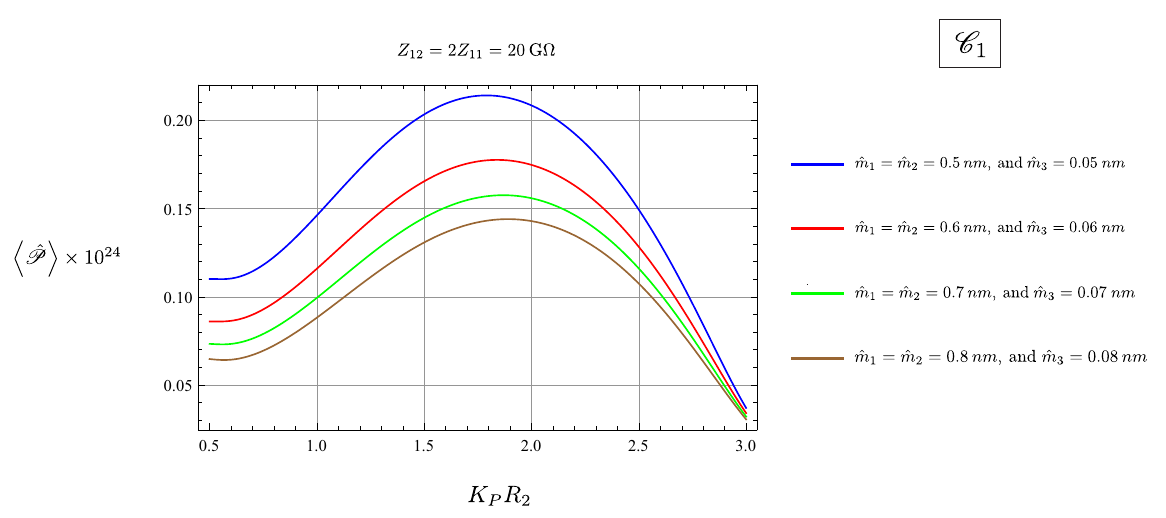}
	\end{center}
	\caption{The effects of the characteristic lengths on the antenna efficiency pertinent to composition $\mathscr{C}_1$.}
	\label{3000}
\end{figure}

\begin{figure}[H]
	\begin{center}
		\includegraphics[width=12cm]{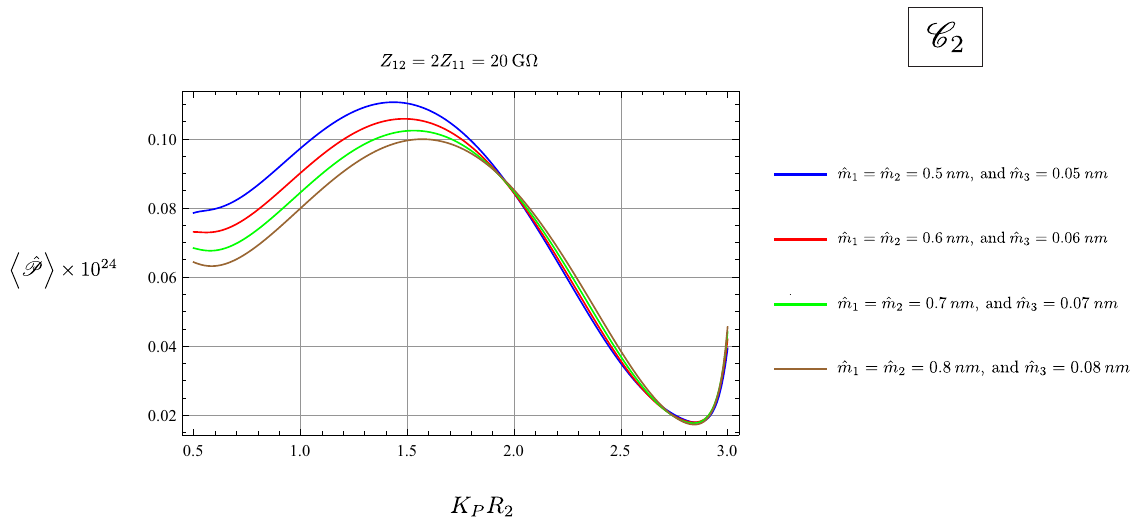}
	\end{center}
	\caption{The effects of the characteristic lengths on the antenna efficiency pertinent to composition $\mathscr{C}_2$.}
	\label{3001}
\end{figure}

\begin{figure}[H]
	\begin{center}
		\includegraphics[width=12cm]{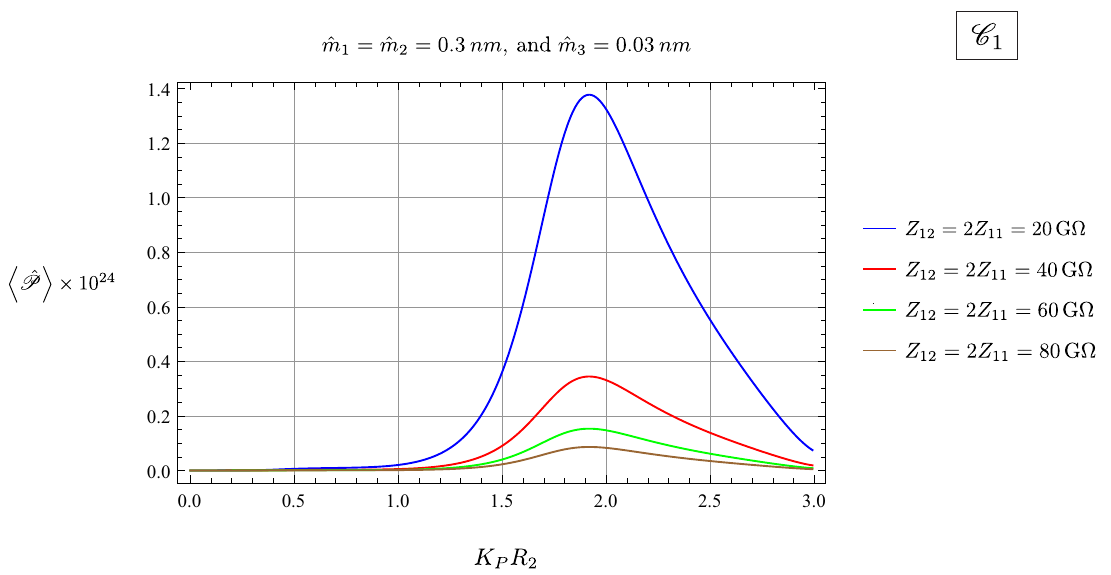}
	\end{center}
	\caption{The effects of the surface and interface impedances on the antenna efficiency pertinent to composition $\mathscr{C}_1$.}
	\label{3002}
\end{figure}

\begin{figure}[H]
	\begin{center}
		\includegraphics[width=12cm]{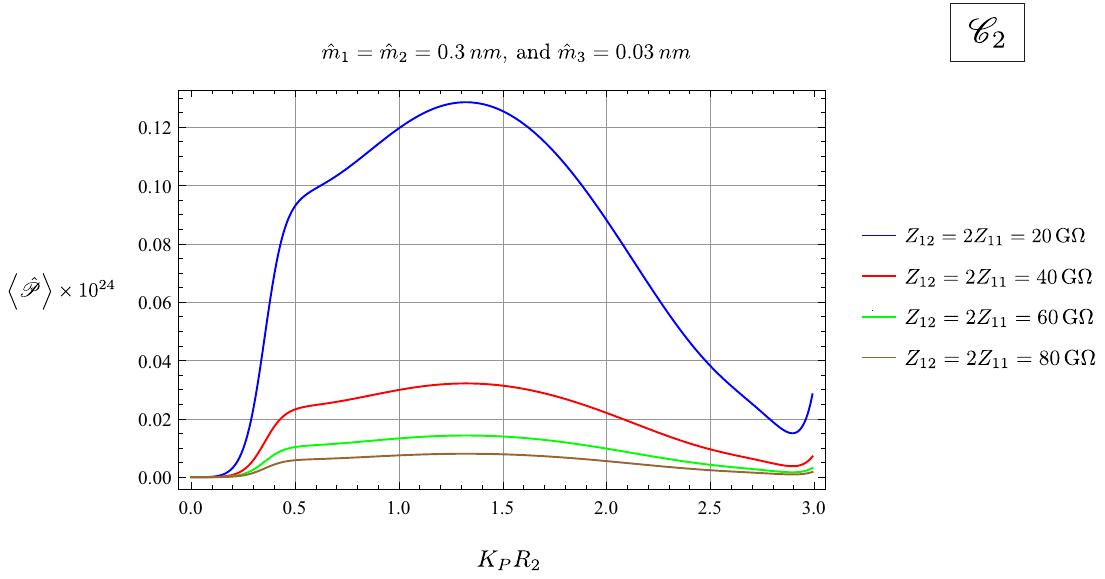}
	\end{center}
	\caption{The effects of the surface and interface impedances on the antenna efficiency pertinent to composition $\mathscr{C}_2$.}
	\label{3003}
\end{figure}

For the free surface and the matrix-shell interface we assume that ${Z_{12}} = 2{Z_{11}} = 20G\Omega$. The effects of the MEE characteristic lengths on the variations of the antenna efficiency, $\left\langle {\mathscr{\hat P}} \right\rangle$ as a function of the normalized incident wavenumber are examined in Figs.~\ref{3000} and \ref{3001} for the MEE shell compositions $\mathscr{C}_1$($\mathrm{40\% BaTi{O_3}}$-$\mathrm{60\% Co{Fe_2}{O_4}}$) and $\mathscr{C}_2$($\mathrm{60\% BaTi{O_3}}$-$\mathrm{40\% Co{Fe_2}{O_4}}$), respectively. As shown in these figures, the frequency of the fundamental mode resonance increases with characteristic length for both compositions, $\mathscr{C}_1$ and $\mathscr{C}_2$. However, the antenna efficiency at this resonance decreases with increasing characteristic length in both cases. Furthermore, the half-power bandwidth decreases with increasing length for 
$\mathscr{C}_1$ while it increases for $\mathscr{C}_2$. A detailed comparison of the figures reveals that, for each characteristic length, the fundamental resonance frequency and antenna efficiency are consistently higher for composition $\mathscr{C}_1$ than for $\mathscr{C}_2$.
On the other hand, for ${{\hat m}_1} = {{\hat m}_2} = 0.3nm$ and ${{\hat m}_3}=0.03 nm$, the effects of the MEE surface and interface impedances on the variations of the antenna efficiency, $\left\langle {\mathscr{\hat P}} \right\rangle$ as a function of the normalized incident wavenumber are addressed in Figs.~\ref{3002} and \ref{3003} for the MEE shell compositins $\mathscr{C}_1$ and $\mathscr{C}_2$, respectively. The results presented in these figures indicate that the frequency of the fundamental mode resonance remains approximately constant across all impedances for both compositions. However, the antenna efficiency at this resonance decreases as impedance increases in both cases. Additionally, the half-power bandwidth increases with impedance for both $\mathscr{C}_1$ and $\mathscr{C}_2$. Further analysis reveals that, for each impedance, the fundamental resonance frequency and antenna efficiency are consistently higher for composition $\mathscr{C}_1$ compared to $\mathscr{C}_2$.

\subsubsection{Case (b) - MEE fields of a nano-sized antenna embedded in a metal matrix}
This example illustrates the occurrence of the plasmonics phenomenon. It is important to note that, in the plasmonic case, it is crucial that the frequency of the scattered electromagnetic wave exceeds the plasma frequency of the host metal matrix to enable radiation. This requirement arises from the relationships outlined in  Eqs.~\eqref{eq54}, \eqref{eq55}, \eqref{eq59}, and \eqref{eq59} along with their asymptotic relations at infinity (Eqs.~\eqref{eq145}-\eqref{eq148}). Specifically, the electromagnetic wavenumber (${K_m}$) and, consequently, the dielectric function (Eq.~\eqref{eq234}) must be positive to ensure non-evanescent behavior. In this case, we examine a nano-sized MEE shell with an outer-to-inner radius ratio of $\frac{{{R_2}}}{{{R_1}}} = 2$, embedded in an aluminum matrix, where the outer radius of the shell is taken as $R_2 =1 \mathrm{nm}$.

\begin{figure}[H]
	\begin{center}
		\includegraphics[width=12cm]{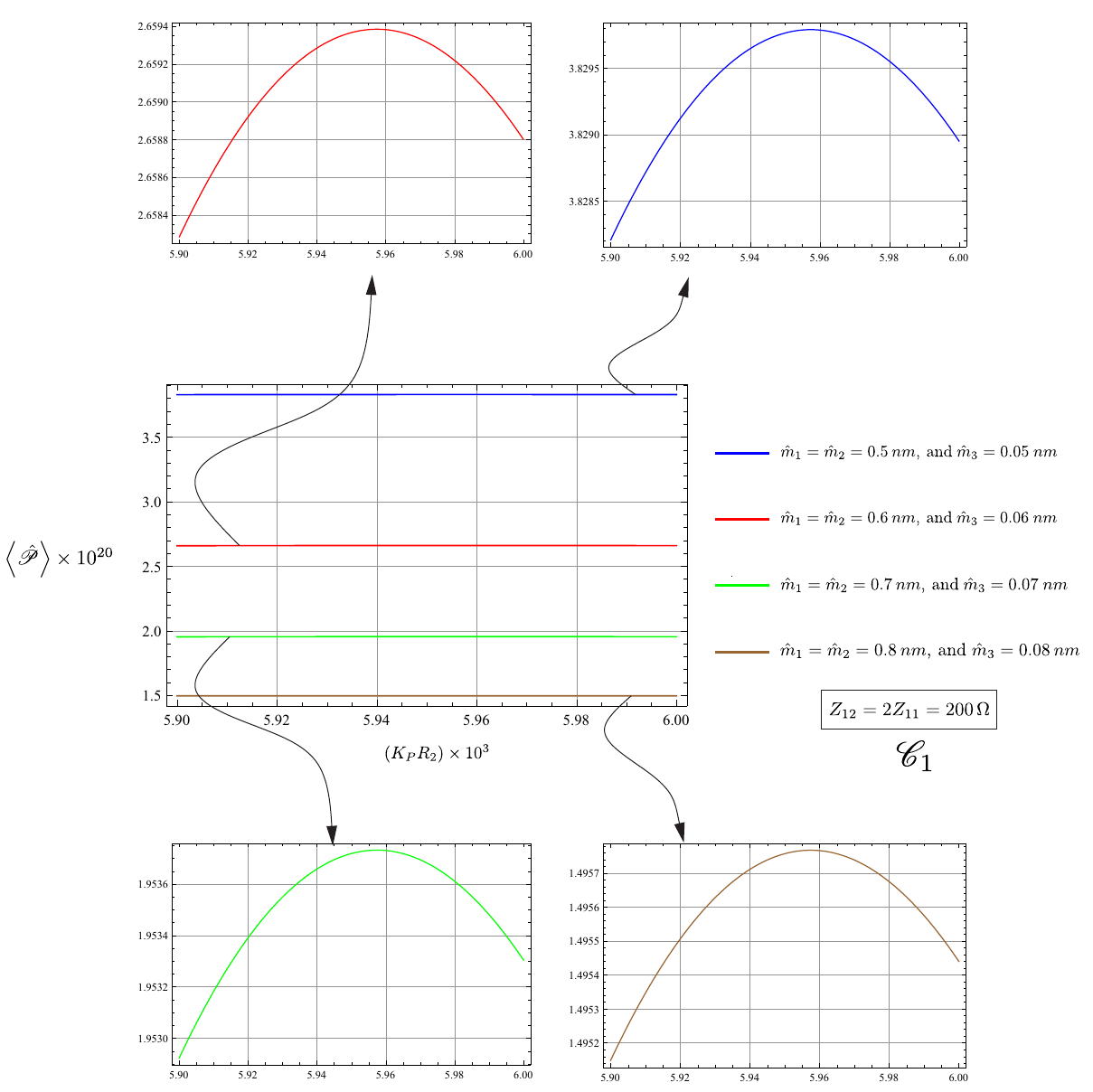}
	\end{center}
	\caption{The effects of the characteristic lengths on the antenna efficiency pertinent to composition $\mathscr{C}_1$.}
	\label{3004}
\end{figure}

\begin{figure}[H]
	\begin{center}
		\includegraphics[width=12cm]{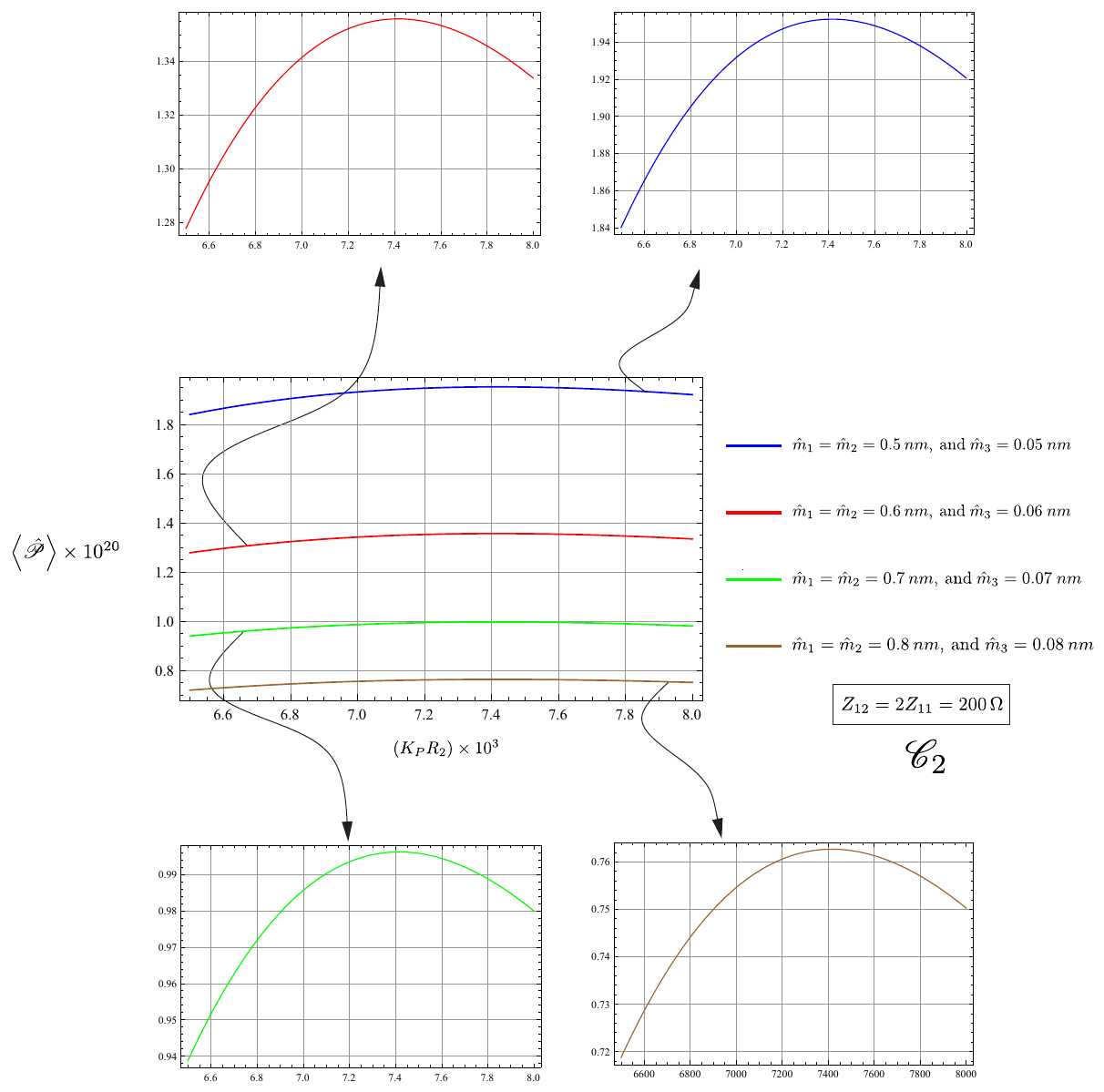}
	\end{center}
	\caption{The effects of the characteristic lengths on the antenna efficiency pertinent to composition $\mathscr{C}_2$.}
	\label{3005}
\end{figure}

\begin{figure}[H]
	\begin{center}
		\includegraphics[width=12cm]{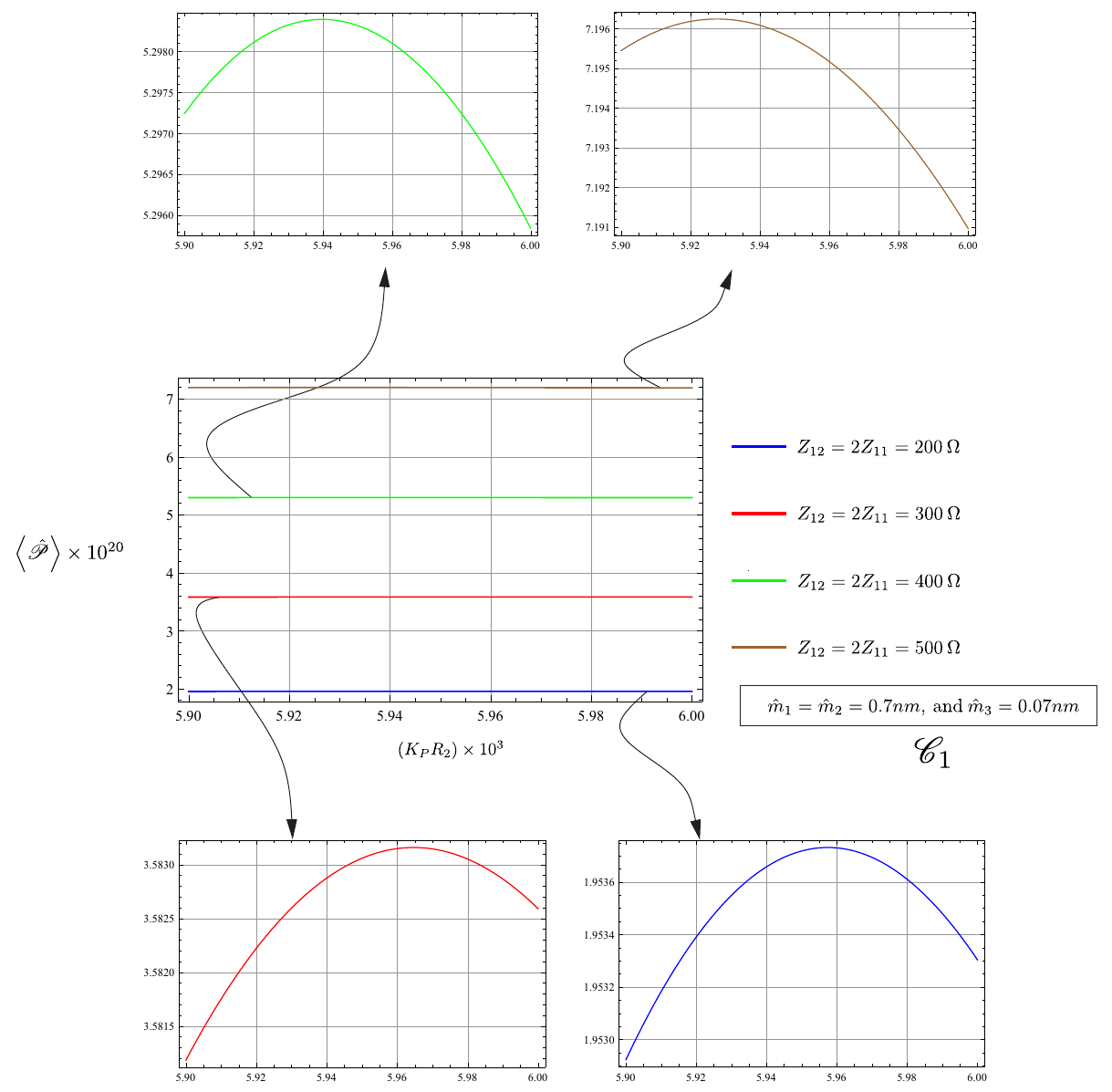}
	\end{center}
	\caption{The effects of the surface and interface impedances on the antenna efficiency pertinent to composition $\mathscr{C}_1$.}
	\label{3006}
\end{figure}

\begin{figure}[H]
	\begin{center}
		\includegraphics[width=12cm]{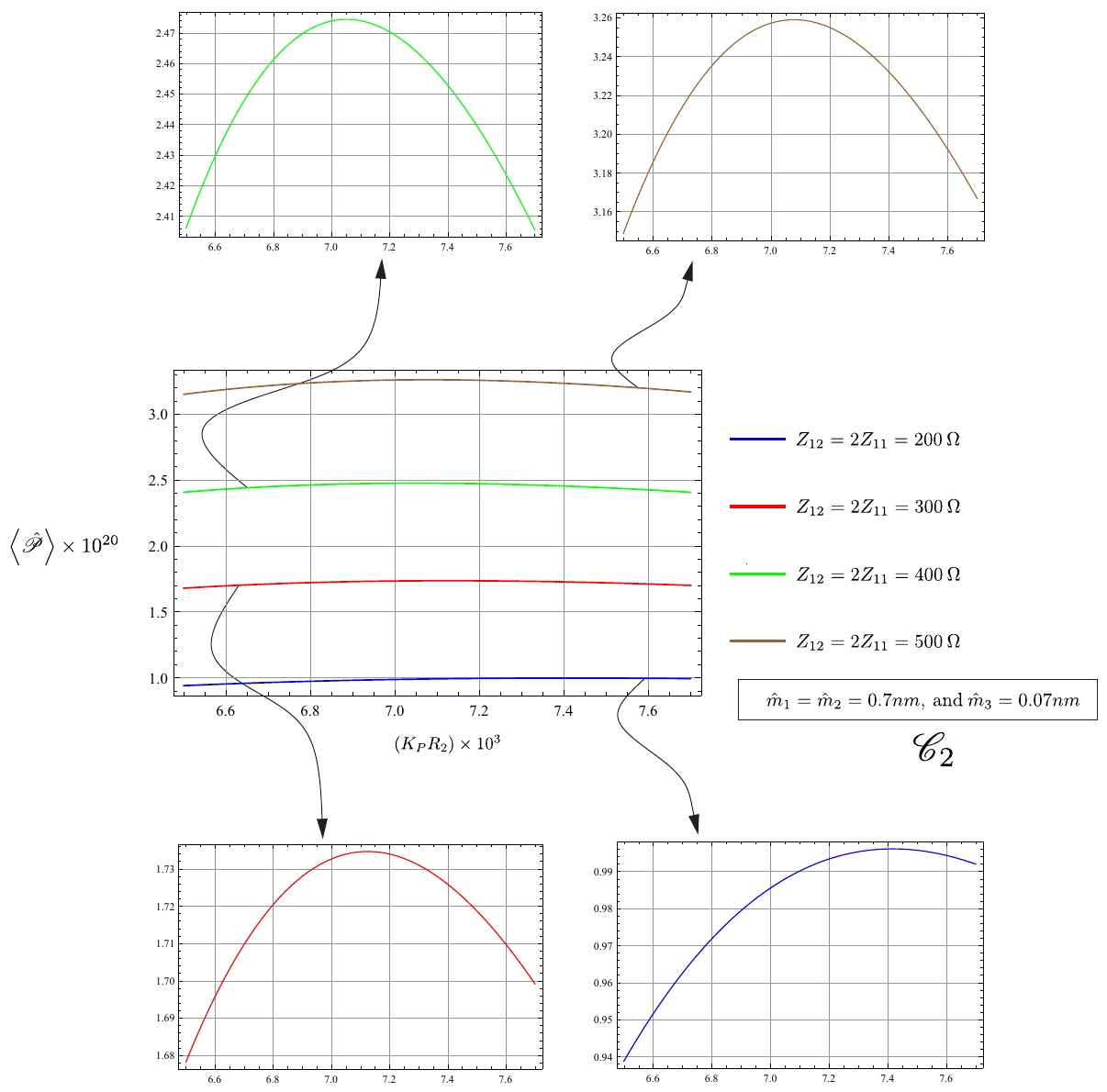}
	\end{center}
	\caption{The effects of the surface and interface impedances on the antenna efficiency pertinent to composition $\mathscr{C}_2$.}
	\label{3007}
\end{figure}

Assuming surface and interface impedances ${Z_{12}} = 2{Z_{11}} = 200\Omega$, we investigate the influence of MEE characteristic lengths on the antenna efficiency, $\left\langle {\mathscr{\hat P}} \right\rangle$, as a function of the normalized incident wavenumber, ${K_p}{R_2}$. Figs.~\ref{3004} and \ref{3005}  illustrate these effects for two MEE shell compositions, $\mathscr{C}_1$($\mathrm{40\% BaTi{O_3}}$-$\mathrm{60\% Co{Fe_2}{O_4}}$) and $\mathscr{C}_2$($\mathrm{60\% BaTi{O_3}}$-$\mathrm{40\% Co{Fe_2}{O_4}}$), respectively. These figures reveal that the frequency of the fundamental mode resonance remains nearly unchanged across all characteristic lengths for both $\mathscr{C}_1$ and $\mathscr{C}_2$. However, the antenna efficiency at resonance lowers down with increasing characteristic length for each composition. Furthermore, a comparison between the two compositions shows that, for a given set of characteristic lengths the composition, $\mathscr{C}_1$ achieves a higher antenna efficiency than $\mathscr{C}_2$, while the resonance frequency is lower for $\mathscr{C}_1$ compared to $\mathscr{C}_2$.
From a different perspective, Figs.~\ref{3006} and \ref{3007} examine the impact of varying MEE surface and interface impedances on antenna efficiency for both compositions, using characteristic lengths ${{\hat m}_1} = {{\hat m}_2} = 0.7 \ \mathrm{nm}$ and ${{\hat m}_2} = 0.07 \ \mathrm{nm}$. These figures indicate that the antenna efficiency at resonance increases with surface and interface impedances in both compositions. A comparative analysis indicates that, for a given surface and interface impedances, composition $\mathscr{C}_1$ consistently exhibits higher antenna efficiency and a lower fundamental resonance frequency than composition $\mathscr{C}_2$.

It should be emphasized that the plasmonic effect is clearly evident in Fig.~\ref{3004} through Fig.~\ref{3007}.  In all of these figures the fundamental resonance occurs at a frequency higher than the plasma frequency of aluminium ($2.24 \times {10^{16}}$ Hz), as expected. It is noteworthy to mention that the  frequency observed in the current example is $3$ order of magnitude higher than the fundamental resonance frequency occurred in the case of the polymer matrix considered in examples \ref{sec10} and \ref{sec1a}. 
\section{Conclusion}
The coupled multiphysics magneto-electro-elastic (MEE) properties of the surface and interface of an embedded nano-sized MEE object struck by incident acoustic waves have a substantial influence on its overall multifunctional behavior. The fact that the interatomic bond lengths, charge density distribution, magnetic moment, and atomic configurations associated with atoms located on the free surface of a nano-sized MEE particle and at the interface between the particle and its surrounding matrix are significantly different from those associated with atoms within the underlying bulk region of the particle suggests that the MEE bulk region, its bounding surface, and interface should be treated as separate entities. In this work, these regions are described by three distinct MEE constitutive relations. Moreover, the reflected and refracted MEE waves, resulting from the interaction of the particle with incident acoustic waves, are governed by three different sets of coupled elastodynamic and Maxwell's equations.

For a precise account of the MEE surface effects, an accurate analysis that combines surface elasticity theory and the equivalent impedance matrix (EIM) method is given in this work for the first time. In this approach, the ME surface/interface is represented by the equivalent T-circuit, which gives rise to the surface/interface impedances; this parameter accommodates both the surface polarization and magnetization. In this framework, the coupled equations describing the MEE surface/interface conditions consist of three relations involving the second derivatives of the spectral displacements and three relations involving the ME fields in the vicinity of the surface and interface of the MEE particle.

To showcase the robustness of the current analysis, several descriptive examples were presented and solved. Moreover, the analysis was also simplified to treat a simpler scenario in which the embedded particle is made of a piezoelectric material \cite{f50} rather than the more general case of the MEE particle considered in the present work. Subsequently, the special case of the PZT-4 shell embedded in an epoxy matrix with negligibly small surface/interface magnetization, which was recently considered in \cite{f50}, has been reconsidered herein for verification purposes; it turned out that the variations in antenna efficiency as a function of the normalized incident wavenumber for various surface/interface characteristic lengths, as calculated herein, are in exact agreement with those obtained in \cite{f50}.

Beyond the verification example, for the MEE shells made of two different compositions,  $\mathscr{C}_1$ (40\% BaTiO$_3$ - 60\% CoFe$_2$O$_4$) and  $\mathscr{C}_2$ (60\% BaTiO$_3$ - 40\% CoFe$_2$O$_4$), embedded in either an epoxy matrix or an aluminum matrix, the impact of the surface/interface intrinsic length scales and impedances was examined in several examples.

In the Results and Discussion section, we illustrated how the half-power bandwidth can be precisely controlled by adjusting the surface/interface characteristic lengths and impedances. By tuning these parameters, the bandwidth can be expanded to accommodate broadband or multi-frequency applications, such as communication systems and Wi-Fi, or narrowed to enhance performance at a specific frequency, making it ideal for high-precision narrowband applications.

\appendix
\section{Simplified EIM method for piezoelectric particles with nonmagnetic surface/interface embedded in an isotropic matrix}\label{App new}
In this Appendix, we will show that in the specific case where the ultra-thin shell is made of a piezoelectric material and embedded in an isotropic matrix with no magnetic properties and surface/interface magnetization,then the expressions for the impedences take on a simple form, ${Z_{11}}={Z_{12}}=\frac{\iota }{{\omega {\kappa ^\wp }}}$. This can readily proved as discussed below.  In the absence of the surface and interface magnetization. In the absence of the surface and interface magnetization (${M^\wp }=0$), it follows from Eqs.~\eqref{eq415} and \eqref{eq416} that:
\begin{subequations}
	\begin{align}
	&\mathbf{n} \times [\mathbf{E}]= 0 \quad\hspace{0.19em}\hspace{0.25em}\hspace{0.25em} \hfil \text{on}\hspace{0.25em}r={R_1},\hspace{0.25em} \text{and}\hspace{0.25em}r={R_2}\label{eq4150}\\
	&\mathbf{n} \times [\mathbf{H}] + \iota \omega {\mathbf{P}^\wp} = 0  \qquad\hspace{0.25em}\hspace{0.25em}\hspace{0.25em} \text{on}\hspace{0.25em}r={R_1},\hspace{0.25em} \text{and}\hspace{0.25em}r={R_2} \label{eq4160}
	\end{align}
\end{subequations}
It should be recalled that Sec~\ref{2-1} was devoted to the modeling of surfaces and interfaces where surface and interface magnetization are non-negligible. As discussed earlier, in general, there are jumps in the magnetoelectric fields across the surfaces and interfaces, regardless of the surface and interface magnetization. Thus, in a manner similar to the modeling of the magnetoelectric behavior of the surfaces and interfaces discussed in Sec~\ref{2-1}, the magnetoelectric behavior of the surfaces and interfaces can be modeled by drawing parallels between the propagation of plane waves and the transmission of signals in a transmission line. In the simple case where the surface and interface magnetization are absent, the pertinent transmission line may be represented as below:
\begin{figure}[H]
	\begin{center}
		\includegraphics[width=10cm]{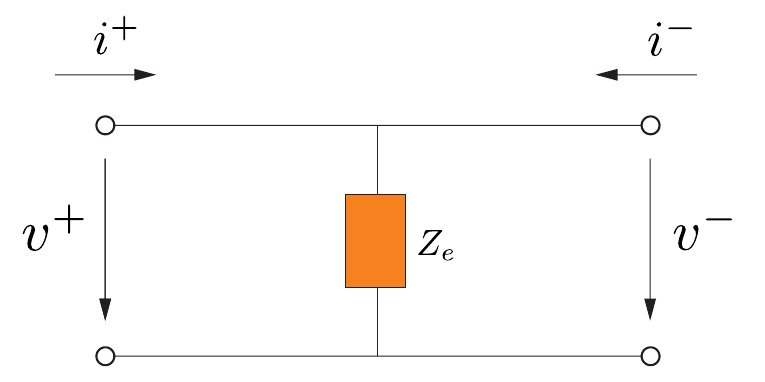}
	\end{center}
	\caption{The equivalent T circuit for the surface/interface ME boundary conditions.}
	\label{300}
\end{figure}
In this transmission line representation, the surface/interface is modeled by a shunt impedance, $Z_e$. The voltages across the impedance remain equal, while the currents experience a discontinuity:
\begin{equation}
{v^ + } - {v^ - } = 0,\quad {i^ + }- {i^ - } = \frac{{{v^ + }}}{{{Z_e}}}
\end{equation}
Therefore, the effective tangential magnetoelectric (ME) fields on both sides of the surface/interface boundaries of the embedded spherical MEE particle can be expressed as:
\begin{equation}\label{eq129}
\mathbf{E}_t^ +  - \mathbf{E}_t^ -  = 0,\quad \mathbf{n} \times [\mathbf{H}_t^ +  - \mathbf{H}_t^ - ] =\frac{{\mathbf{E}_t^ + }}{{{Z_e}}}
\end{equation}
Now, comparing Eq.~\eqref{eq129} with the previously mentioned equations, $\mathbf{n} \times [{\mathbf{H}_t^ +}-{\mathbf{H}t^ -} ]={\mathbf{J}_{tot,e}}=- \iota \omega {\mathbf{P}^\wp }$ and ${\mathbf{E}_t^ +}-{\mathbf{E}_t^ -} =0$, and applying the constitutive equations of the surface/interface (Eq.\eqref{eq371} and \eqref{eq372}), we derive the following result. (Notably, in this scenario, there is no magnetoelectric effect ($\mathscr{G}_{11} = 0$), and residual surface/interface electric polarization ($\mathbf{P}_0$) is neglected)
\begin{equation}
{Z_e}=\frac{\iota }{{\omega {\kappa ^\wp }}}
\end{equation}
If we expand the vectorial form of Eq.~\eqref{eq129}, the relevant boundary conditions will be analogous to those in Eqs.~\eqref{eq422} and \eqref{eq423}; however, this time we have ${Z_{11}}={Z_{12}}={Z_e}=\frac{\iota }{{\omega {\kappa ^\wp }}}$.
\section{Components of the tensors $\mathbf{Q}_i$}
Here, we provide explicit expressions for the components of the tensors $Q_i$ ($i=0,1,2,3$) appearing in Eq.~\eqref{eq110}
\begin{equation}
{\mathbf{Q}_{i}} = {q_{ijk}}{\mathbf{e}_j} \otimes {\mathbf{e}_k},
\end{equation}
in which $j,k=1,2,3$. We have
\begin{align}
&{q_{011}} = {C_{33}} + \frac{{{\gamma _{33}}\big(2{d_{33}}{e_{33}} - {g_{33}}{\gamma _{33}}\big) - {\mu _{33}}e_{33}^2}}{{d_{33}^2 - {g_{33}}{\mu _{33}}}},\quad{q_{012}} = {q_{013}} = {q_{014}} = 0,\notag\\
&{q_{111}} = \frac{{2\bigg({C_{33}}d_{33}^2 + 2{d_{33}}{e_{33}}{\gamma _{33}} - {g_{33}}\gamma _{33}^2 - {\mu _{33}}\big(e_{33}^2 + {C_{33}}{g_{33}}\big)\bigg)}}{{d_{33}^2 - {g_{33}}{\mu _{33}}}},\notag\\
&{q_{112}} =- l(l + 1)\big({C_{13}} + {C_{44}}\big) + \frac{{l(l + 1)\big({d_{33}}{e_{31}} - {g_{33}}{\gamma _{31}}\big)\big({d_{33}}{e_{33}} - {g_{33}}{\gamma _{33}}\big)}}{{{g_{33}}\big(d_{33}^2 - {g_{33}}{\mu _{33}}\big)}}\notag\\
&\quad\quad\quad - \frac{{l(l + 1){d_{11}}\big({d_{11}}{e_{31}}{e_{33}} + 2{g_{33}}{e_{15}}{\gamma _{15}}\big)}}{{{g_{33}}\big(d_{11}^2 - {g_{11}}{\mu _{11}}\big)}}+ \frac{{l(l + 1)\big({\mu _{11}}e_{15}^2 + {g_{11}}\gamma _{15}^2\big)}}{{\big(d_{11}^2 - {g_{11}}{\mu _{11}}\big)}}\notag\\
& \quad\quad\quad+ \frac{{l(l + 1){g_{11}}{\mu _{11}}{e_{31}}{e_{33}}}}{{{g_{33}}\big(d_{11}^2 - {g_{11}}{\mu _{11}}\big)}},\notag\\
&{q_{113}} = \frac{{l(l + 1)\iota }}{\omega }\bigg(\frac{{{d_{11}}{e_{15}} - {g_{11}}{\gamma _{15}}}}{{d_{11}^2 - {g_{11}}{\mu _{11}}}} - \frac{{{d_{33}}{e_{33}} - {g_{33}}{\gamma _{33}}}}{{d_{33}^2 - {g_{33}}{\mu _{33}}}}\bigg),\notag\\
&{q_{114}} = - \frac{{l(l + 1)\iota }}{\omega }\bigg(\frac{{{d_{11}}{\gamma _{15}}}}{{d_{11}^2 - {g_{11}}{\mu _{11}}}}\bigg) - \frac{{l(l + 1)\iota }}{\omega }\bigg(\frac{{d_{11}^2({\mu _{33}}{e_{33}} - {d_{33}}{\gamma _{33}})}}{{\big(d_{11}^2 - {g_{11}}{\mu _{11}}\big)\big(d_{33}^2 - {g_{33}}{\mu _{33}}\big)}}\bigg)\notag\\
& \quad\quad\quad+ \frac{{l(l + 1)\iota {\mu _{11}}}}{\omega }\bigg(\frac{{{\mu _{33}}{g_{11}}{e_{33}} - {d_{33}}{g_{11}}{\gamma _{33}}}}{{\big(d_{11}^2 - {g_{11}}{\mu _{11}}\big)\big(d_{33}^2 - {g_{33}}{\mu _{33}}\big)}}\bigg)+ \frac{{l(l + 1)\iota }}{\omega }\bigg(\frac{{{\mu _{11}}{e_{15}}}}{{d_{11}^2 - {g_{11}}{\mu _{11}}}}\bigg),\notag\\
&{q_{311}} =2{C_{13}} + \frac{{2\big({d_{33}}{e_{31}} - {g_{33}}{\gamma _{31}}\big)\bigg({d_{33}}\big(2{e_{31}} - {e_{33}}\big) + {g_{33}}\big({\gamma _{33}} - 2{\gamma _{31}}\big)\bigg)}}{{{g_{33}}\big(d_{33}^2 - {g_{33}}{\mu _{33}}\big)}}\notag\\
&\quad\quad\quad + \frac{{d_{11}^2\bigg(2{e_{31}}\big({e_{33}} - 2{e_{31}}\big) - 2\big({C_{11}} + {C_{11}}\big){g_{33}} - l(l + 1){C_{44}}{g_{33}}\bigg)}}{{{g_{33}}\big(d_{11}^2 - {g_{11}}{\mu _{11}}\big)}}\notag\\
& \quad\quad\quad- \frac{{  l(l + 1)\bigg(2{g_{33}}{d_{11}}{e_{15}}{\gamma _{15}} - {g_{11}}{g_{33}}\gamma _{15}^2 - {g_{33}}{\mu _{11}}e_{15}^2 - {C_{44}}{g_{11}}{g_{33}}{\mu _{11}}\bigg)}}{{{g_{33}}\big(d_{11}^2 - {g_{11}}{\mu _{11}}\big)}}\notag\\
&\quad\quad\quad +\frac{{{g_{11}}{\mu _{11}}\bigg(4e_{31}^2 - 2{e_{31}}{e_{33}} + 2\big({C_{11}} + {C_{12}}\big)\bigg)}}{{{g_{33}}\big(d_{11}^2 - {g_{11}}{\mu _{11}}\big)}},\notag\\
&{q_{312}} = - {C_{13}} + \frac{{2\big({d_{33}}{e_{11}} - {g_{33}}{\gamma _{31}}\big)\bigg({d_{33}}\big(2{e_{31}} - {e_{33}}\big) + {g_{33}}\big({\gamma _{33}} - 2{\gamma _{31}}\big)\bigg)}}{{{g_{33}}\big(d_{33}^2 - {g_{33}}{\mu _{33}}\big)}}\notag\\
& \quad\quad\quad- \frac{{{\mu _{11}}\bigg(2e_{31}^2{g_{11}} - {e_{31}}{e_{33}}{g_{11}} + e_{15}^2{g_{33}} + \big({C_{11}} + {C_{12}} + {C_{44}}\big){g_{11}}{g_{33}}\bigg)}}{{{g_{33}}\big(d_{11}^2 - {g_{11}}{\mu _{11}}\big)}}\notag\\
& \quad\quad\quad+ \frac{{d_{11}^2\bigg(2e_{31}^2 - {e_{31}}{e_{33}} + \big({C_{11}} + {C_{12}} + {C_{44}}\big){g_{33}}\bigg)}}{{{g_{33}}\big(d_{11}^2 - {g_{11}}{\mu _{11}}\big)}}\notag\\
&\quad\quad\quad+ \frac{{{\gamma _{15}}\big(2{d_{11}}{e_{15}} - {g_{11}}{g_{33}}{\gamma _{15}}\big)}}{{d_{11}^2 - {g_{11}}{\mu _{11}}}},\notag\\
&{q_{313}} = \frac{{l(l + 1)\iota }}{\omega }\bigg(\frac{{{d_{11}}{e_{15}} - {g_{11}}{\gamma _{15}}}}{{d_{11}^2 - {g_{11}}{\mu _{11}}}} + \frac{{2\big({d_{33}}{e_{31}} - {g_{33}}{\gamma _{31}}\big) + {g_{33}}{\gamma _{33}} - {d_{33}}{e_{33}}}}{{d_{33}^2 - {g_{33}}{\mu _{33}}}}\bigg),\notag\\
&{q_{314}} =\frac{{l(l + 1)\iota }}{{\omega {g_{33}}}}\bigg(\frac{{{\mu _{11}}{g_{33}}{e_{15}} + {\mu _{11}}{g_{11}}\big(2{e_{31}} - {e_{33}}\big)}}{{d_{11}^2 - {g_{11}}{\mu _{11}}}}\bigg)\notag\\
&\quad\quad\quad - \frac{{l(l + 1)\iota }}{{\omega {g_{33}}}}\bigg(\frac{{d_{11}^2\big(2{e_{31}} - {e_{33}}\big) + {d_{11}}{g_{33}}{\gamma _{15}}}}{{d_{11}^2 - {g_{11}}{\mu _{11}}}}\bigg)\notag\\
&\quad\quad\quad + \frac{{l(l + 1)\iota }}{{\omega {g_{33}}}}\bigg(\frac{{d_{33}^2\big(2{e_{31}} - {e_{33}}\big) - {d_{33}}{g_{33}}\big(2{\gamma _{31}} - {\gamma _{33}}\big)}}{{d_{33}^2 - {g_{33}}{\mu _{33}}}}\bigg),\notag\\
&{q_{411}} = {q_{412}} = 0,\quad{q_{413}} = {e_{15}},\quad{q_{414}} = {\gamma _{15}},\notag\\
&{q_{511}} = \rho {\omega ^2},\quad{q_{512}} = {q_{513}} = {q_{514}} = 0,\notag\\
&{q_{021}} = 0,\quad{q_{022}} = {C_{44}} + \frac{{{\gamma _{15}}(2{d_{11}}{e_{15}} - {g_{11}}{\gamma _{15}}) - {\mu _{11}}e_{15}^2}}{{d_{11}^2 - {g_{11}}{\mu _{11}}}},\notag\\
&{q_{023}} =  - \frac{{\iota ({d_{11}}{e_{15}} - {g_{11}}{\gamma _{15}})}}{{\omega (d_{11}^2 - {g_{11}}{\mu _{11}})}},\quad{q_{024}} =  - \frac{{\iota ({\mu _{11}}{e_{15}} - {d_{11}}{\gamma _{15}})}}{{\omega (d_{11}^2 - {g_{11}}{\mu _{11}})}},\notag\\
&{q_{121}} = {C_{13}} + {C_{44}} + \frac{{2{d_{11}}{e_{15}}{\gamma _{15}}}}{{d_{11}^2 - {g_{11}}{\mu _{11}}}} - \frac{{{g_{11}}\gamma _{15}^2}}{{d_{11}^2 - {g_{11}}{\mu _{11}}}} - \frac{{{\mu _{11}}e_{15}^2}}{{d_{11}^2 - {g_{11}}{\mu _{11}}}}\notag\\
& \quad\quad\quad- \frac{{{g_{33}}{\gamma _{31}}{\gamma _{33}} - {d_{33}}({e_{33}}{\gamma _{31}} + {e_{31}}{\gamma _{33}}) + {e_{31}}{e_{33}}{\mu _{33}}}}{{d_{33}^2 - {g_{33}}{\mu _{33}}}},\notag\\
&{q_{122}} = 2{C_{44}} + 2\frac{{\bigg({\gamma _{15}}\big(2{d_{11}}{e_{15}} - {g_{11}}{\gamma _{15}}\big) - {\mu _{11}}e_{15}^2\bigg)}}{{d_{11}^2 - {g_{11}}{\mu _{11}}}},\notag\\
&{q_{123}} =  - 4\frac{{\iota \big({d_{11}}{e_{15}} - {g_{11}}{\gamma _{15}}\big)}}{{\omega \big(d_{11}^2 - {g_{11}}{\mu _{11}}\big)}},\quad{q_{124}} =  - 4\frac{{\iota ({\mu _{11}}{e_{15}} - {d_{11}}{\gamma _{15}})}}{{\omega (d_{11}^2 - {g_{11}}{\mu _{11}})}},\notag\\
&{q_{221}} = {q_{222}} = 0,\quad{q_{223}} =  - {e_{15}},\quad{q_{224}} =  - {\gamma _{15}},\notag\\
&{q_{321}} = 2{C_{44}} - 2\frac{{{{\big({d_{33}}{e_{31}} - {g_{33}}{\gamma _{31}}\big)}^2}}}{{d_{33}^2 - {g_{33}}{\mu _{33}}}} + \frac{{d_{11}^2\bigg(2e_{31}^2 + \big({C_{11}} + {C_{12}}\big){g_{33}}\bigg)}}{{{g_{33}}\big(d_{11}^2 - {g_{11}}{\mu _{11}}\big)}}\notag\\
&\quad\quad\quad- \frac{{{\mu _{11}}\bigg(2{g_{11}}e_{31}^2 + {g_{33}}\big(2e_{15}^2 + ({C_{11}} + {C_{12}}){g_{11}}\big)\bigg)}}{{{g_{33}}\big(d_{11}^2 - {g_{11}}{\mu _{11}}\big)}}\notag \\
&\quad\quad\quad + \frac{{2{\gamma _{15}}\big(2{d_{11}}{e_{15}} - {g_{11}}{\gamma _{15}}\big)}}{{d_{11}^2 - {g_{11}}{\mu _{11}}}},\notag\\
&{q_{322}} = - \big({l^2} + l - 1\big){C_{11}} + \frac{{l(l + 1){{\big({d_{33}}{e_{31}} - {g_{33}}{\gamma _{31}}\big)}^2}}}{{{g_{33}}\big(d_{11}^2 - {g_{11}}{\mu _{11}}\big)}}\notag\\
& \quad\quad\quad- \frac{{d_{11}^2\big(l(l + 1)e_{31}^2 + 2{C_{44}}{g_{33}}\big)}}{{{g_{33}}\big(d_{11}^2 - {g_{11}}{\mu _{11}}\big)}} - \frac{{2{\gamma _{15}}\big(2{d_{11}}{e_{15}} - {g_{11}}{\gamma _{15}}\big)}}{{d_{11}^2 - {g_{11}}{\mu _{11}}}}\notag\\
& \quad\quad\quad+ \frac{{l(l + 1){g_{11}}{\mu _{11}}e_{31}^2 + 2{g_{33}}{\mu _{11}}e_{15}^2 + ({C_{12}} + 2{C_{44}}){g_{11}}{g_{33}}{\mu _{11}}}}{{{g_{33}}(d_{11}^2 - {g_{11}}{\mu _{11}})}},\notag\\
&{q_{323}} = \frac{{2\iota }}{\omega }(\frac{{{g_{11}}{\gamma _{15}} - {e_{15}}{d_{11}}}}{{d_{11}^2 - {g_{11}}{\mu _{11}}}}) + \frac{{l(l + 1)\iota }}{\omega }(\frac{{{g_{33}}{\gamma _{31}} - {d_{33}}{e_{31}}}}{{d_{33}^2 - {g_{33}}{\mu _{33}}}}),\notag\\
&{q_{324}} = \frac{{2\iota }}{\omega }(\frac{{{d_{11}}{\gamma _{15}} - {\mu _{11}}{e_{15}}}}{{d_{11}^2 - {g_{11}}{\mu _{11}}}}) - \frac{{l(l + 1)\iota }}{\omega }(\frac{{{\mu _{33}}{e_{31}} - {d_{33}}{\gamma _{31}}}}{{d_{33}^2 - {g_{33}}{\mu _{33}}}}),\notag\\
&{q_{421}} = {q_{422}} = 0,\quad{q_{423}} =  - 3{e_{15}},\quad{q_{424}} =  - 3{\gamma _{15}},\notag\\
&{q_{521}} = {q_{523}} = {q_{524}} = 0,\quad{q_{522}} = \rho {\omega ^2},\notag\\
&{q_{031}} = 0,\quad{q_{032}} = \frac{{{\mu _{11}}{e_{15}} - {d_{11}}{\gamma _{15}}}}{{d_{11}^2 - {g_{11}}{\mu _{11}}}},\notag\\
&{q_{033}} = \frac{\iota }{\omega }(\frac{{{d_{11}}}}{{d_{11}^2 - {g_{11}}{\mu _{11}}}}),\quad{q_{034}} = \frac{\iota }{\omega }(\frac{{{\mu _{11}}}}{{d_{11}^2 - {g_{11}}{\mu _{11}}}}),\notag\\
&{q_{131}} = \frac{{{e_{33}}}}{{{g_{33}}}} + \frac{{{\mu _{11}}{e_{15}} - {d_{11}}{\gamma _{15}}}}{{d_{11}^2 - {g_{11}}{\mu _{11}}}} + \frac{{{g_{33}}{d_{33}}{\gamma _{33}} - {e_{33}}d_{33}^2}}{{{g_{33}}(d_{33}^2 - {g_{33}}{\mu _{33}})}},\notag\\
&{q_{132}} = 0,\quad{Q_{133}} = \frac{{2\iota }}{\omega }(\frac{{{d_{11}}}}{{d_{11}^2 - {g_{11}}{\mu _{11}}}})\quad{q_{134}} = \frac{{2\iota }}{\omega }(\frac{{{\mu _{11}}}}{{d_{11}^2 - {g_{11}}{\mu _{11}}}}),\notag\\
&{q_{231}} = {q_{232}} = {q_{233}} = {q_{234}} = 0,\notag\\
&{q_{331}} =  - 2(\frac{{{\mu _{33}}{e_{31}} - {d_{33}}{\gamma _{31}}}}{{d_{33}^2 - {g_{33}}{\mu _{33}}}}),\quad{q_{332}} = l(l + 1)(\frac{{{\mu _{33}}{e_{31}} - {d_{33}}{\gamma _{31}}}}{{d_{33}^2 - {g_{33}}{\mu _{33}}}}),\notag\\
&{q_{333}} =  - \frac{{l(l + 1)\iota }}{\omega }(\frac{{{d_{33}}}}{{d_{33}^2 - {g_{33}}{\mu _{33}}}}),\quad{q_{334}} =  - \frac{{l(l + 1)\iota }}{\omega }(\frac{{{\mu _{33}}}}{{d_{33}^2 - {g_{33}}{\mu _{33}}}})\notag\\
&{q_{431}} = {q_{432}} = {q_{433}} = {q_{434}} = 0,\notag\\
&{q_{531}} = {q_{532}} = 0,\quad{q_{533}} =  - \iota \omega {d_{11}},\quad{q_{534}} =  - \iota \omega {\mu _{11}},\notag\\
&{q_{041}} = 0,\quad{q_{042}} = \frac{{{g_{11}}{\gamma _{15}} - {d_{11}}{e_{15}}}}{{d_{11}^2 - {g_{11}}{\mu _{11}}}},\notag\\
&{q_{043}} =  - \frac{\iota }{\omega }(\frac{{{g_{11}}}}{{d_{11}^2 - {g_{11}}{\mu _{11}}}}),\quad{q_{044}} =  - \frac{\iota }{\omega }(\frac{{{d_{11}}}}{{d_{11}^2 - {g_{11}}{\mu _{11}}}}),\notag\\
&{q_{141}} = \frac{{{g_{11}}{\gamma _{15}} - {d_{11}}{e_{15}}}}{{d_{11}^2 - {g_{11}}{\mu _{11}}}} + \frac{{{d_{33}}{e_{33}} - {g_{33}}{\gamma _{33}}}}{{d_{33}^2 - {g_{33}}{\mu _{33}}}},\quad{q_{142}} = 0,\notag\\
&{q_{143}} =  - \frac{{2\iota }}{\omega }\frac{{{g_{11}}}}{{(d_{11}^2 - {g_{11}}{\mu _{11}})}},\quad{q_{144}} =  - \frac{{2\iota }}{\omega }\frac{{{d_{11}}}}{{(d_{11}^2 - {g_{11}}{\mu _{11}})}},\notag\\
&{q_{241}} = {q_{242}} = {q_{243}} = {q_{244}} = 0,\notag\\
&{q_{341}} = 2\frac{{{d_{33}}{e_{31}} - {g_{33}}{\gamma _{31}}}}{{d_{33}^2 - {g_{33}}{\mu _{33}}}},\quad{q_{342}} = l(l + 1)\frac{{({g_{33}}{\gamma _{31}} - {d_{33}}{e_{31}})}}{{d_{33}^2 - {g_{33}}{\mu _{33}}}},\notag\\
&{q_{343}} = \frac{{l(l + 1)\iota }}{\omega }(\frac{{{g_{33}}}}{{d_{33}^2 - {g_{33}}{\mu _{33}}}}),\quad{q_{344}} = \frac{{l(l + 1)\iota }}{\omega }(\frac{{{d_{33}}}}{{d_{33}^2 - {g_{33}}{\mu _{33}}}}),\notag\\
&{q_{441}} = {q_{442}} = {q_{443}} = {q_{444}} = 0,\notag\\
&{q_{541}} = {q_{542}} = 0,\quad{q_{543}} = \iota \omega {g_{11}},\quad{q_{544}} = \iota \omega {d_{11}}.\notag
\end{align}

\newpage
\bibliographystyle{apa}\biboptions{authoryear}
\bibliography{mohsen}
\end{document}